\documentclass[12pt]{article}
\usepackage{graphicx}
\begin{document}

\begin{center}

{\Large \bf Integration of Dirac's Efforts to construct Lorentz-covariant
Quantum Mechanics}

\vspace{3mm}
Young S. Kim \\
Center for Fundamental Physics, University of Maryland, College
  Park, MD 20742, USA \\
\vspace{3mm}
Marilyn E. Noz \\
Department of Radiology, New York University, New York, NY 10016,
  USA

\end{center}

\vspace{10mm}

\abstract{The lifelong efforts of Paul A. M. Dirac were to construct
localized quantum systems in the Lorentz covariant world.  In 1927,
he noted that the time-energy uncertainty should be included in
the Lorentz-covariant picture.  In 1945, he attempted to construct
a representation of the Lorentz group using a normalizable Gaussian
function localized both in the space and time variables.  In 1949,
he introduced his instant form to exclude time-like oscillations.
He also introduced the light-cone coordinate system for Lorentz boosts.
Also in 1949, he stated the Lie algebra of the inhomogeneous Lorentz
group can serve as the uncertainty relations in the Lorentz-covariant
world.  It is possible to integrate these three papers to produce the
harmonic oscillator wave function which can be Lorentz-transformed.
In addition, Dirac, in 1963, considered two coupled oscillators to
derive the Lie algebra for the generators of the $O(3,\,2)$ de Sitter
group, which~has ten generators. It is proven possible to contract this
group to the inhomogeneous Lorentz group with ten generators,
which~constitute the fundamental symmetry of quantum mechanics in
Einstein's Lorentz-covariant world.}

 \vspace{50mm}

\noindent Published in Symmetry, Vol. 12(8), 1720 (2020).

\newpage

\section{Introduction}

Since 1973~\cite{kn73}, the present authors have been publishing
papers on the harmonic oscillator wave functions which can be
Lorentz-boosted, leading to a number of
books~\cite{knp86,bkn15iop,knp18,bkn19iop}.  We noticed~that
the Gaussian form of the wave function undergoes an elliptic
space-time deformation when Lorentz-boosted, leading to the
correct proton form factor behavior for large momentum
transfers.

It was then noted that the harmonic oscillator functions
exhibit the contraction and orthogonality properties quite consistent
with known rules of special relativity and quantum
mechanics~\cite{ruiz74,kno79ajp}.

In 1977, using the Lorentz-covariant wave functions, the present
authors showed that Gell-Mann's quark model~\cite{gell64} for
the hadron at rest and Feynman's parton model for fast-moving
hadrons~\cite{fey69a} are two different manifestations of one
Lorentz-covariant entity~\cite{kn77par,kim89}.

In 1979~\cite{kno79jmp}, it was shown that the oscillator system we
constructed can be used for a representation of the Lorentz group,
particularly Wigner's $O(3)$-like little group for massive
particles~\cite{wig39}.  More~recently, it was shown that these
oscillator wave functions can serve as squeezed states of light
and Gaussian entanglement~\cite{yuen76,kn05job,bkn16}.

In 1983, it became known that, if the speed of a spin-1 particle
reaches that of light, the component of spin in the direction of
its momentum remains invariant as its helicity, but the spin
components perpendicular to the momentum become one gauge degree of
freedom~\cite{hks83pl,kiwi90jmp}.    Indeed, this result allows us
to include our oscillator-based quark-parton picture as further
content of Einstein's \mbox{$E = mc^2$~\cite{kim89}}, as shown in
Table~\ref{further}.

\begin{table}
\caption{Lorentz covariance of particles both massive and massless.
The little group of Wigner unifies, for massive and massless particles,
the internal space-time symmetries.  The challenge for us is to find
another unification: that which unifies, in the physics of the high-energy
realm, both the quark and parton pictures. In this paper, we achieve this
purpose by integrating Dirac's three papers.  A similar table was
published in Ref.~\cite{kim89}.}\label{further}
\vspace{0.5mm}
\begin{center}
\begin{tabular}{lccc}
\hline
\hline\\[-0.4ex]
{} & \textbf{Massive, Slow} \hspace{8mm} & \textbf{COVARIANCE} \hspace{8mm}&
{\bf Massless, Fast} \\[2mm]
\hline\\[2mm]
Energy- & {}  & Einstein's & {} \\
Momentum & $E = p^{2}/2m$ & $ E = \sqrt{(cp)^{2} + \left(mc^{2}\right)^{2}} $
& $E = cp$
\\[2mm]
\hline \\
Internal & $S_{3}$ & {}  &  $S_{3}$ \\[-1mm]
Space-time &{} & Wigner's  & {Gauge} \\[-1mm]
Symmetry & $S_{1},\, S_{2}$ & Little Groups &  Transformation \\[2mm]
\hline\\
Relativistic & {} & Integration  &  {} \\
Extended & Quark Model & of Dirac's papers  & Parton Model \\[-1mm]
Particles & {} & 1927,\, 1945,\, 1949 & {} \\
\hline
\hline\\[0.4ex]
\end{tabular}
\end{center}
\end{table}

The purpose of the present paper is to show that the harmonic oscillator wave
functions we have studied since 1973 can serve another purpose.  It is
possible to obtain this covariant form of relativistic extended
particles by integrating the three
papers Dirac wrote, as indicated in Table~\ref{further}.

  \begin{itemize}
   \item[1.] In 1927, Dirac pointed out that the time-energy uncertainty
             should be considered if the system is to be
             Lorentz-covariant~\cite{dir27}.

   \item[2.] In 1945, Dirac said the Gaussian form could serve as
             a representation of the Lorentz group~\cite{dir45}.

   \item[3.] In 1949, when Dirac introduced both his instant form of quantum
              mechanics and his light-cone coordinate
             system~\cite{dir49}, he clearly stated
             that finding a representation of the inhomogeneous
             Lorentz group was the task of Lorentz-covariant quantum
             mechanics.

   \item[4.] In 1963, Dirac used the symmetry of two coupled oscillators to
             construct the $O(3,\,2)$ group~\cite{dir63}.

   \end{itemize}

In the fourth paper published in 1963, Dirac considered two coupled
oscillators using step-up and step-down operators.  He then
constructed the Lie algebra (closed set of commutation relations for the generators)
of the de Sitter group, also known as $O(3,\,2)$, using ten quadratic forms of those
step-up and step-down operators.  The harmonic oscillator is a
language of quantum mechanics while the de Sitter group is a language
of the Lorentz group or Einstein's special relativity.  Thus, his
1963 paper~\cite{dir63} provides the first step toward a unified
view of quantum mechanics and special relativity.

In spite of all those impressive aspects, the above-listed papers are
largely unknown in the physics world.  The reason is that there are
soft spots in those papers.  Dirac firmly believed that one can
construct physical theories only by constructing beautiful
mathematics~\cite{dir70}.  His Dirac equation is a case in point.
Indeed, all of his papers are like poems~\cite{farmelo}.  They are
mathematical poems.  However,~there~are things he did not do.

First, his papers do not contain any graphical illustrations.
For instance, his light-cone coordinate system could be illustrated as
a squeeze transformation, but he did not draw a picture~\cite{dir49}.
When he talked about the Gaussian function~\cite{dir45} in the
space-time variables, he could have used a circle in the
two-dimensional space of the longitudinal and time-like variables.

Second, Dirac seldom made reference to his own earlier papers.
In his 1945 paper~\cite{dir45}, there is a distribution along
the time-like variable, but he did not mention his earlier paper
of 1927~\cite{dir27} where the time-energy uncertainty was discussed.
In his 1949 paper, when he proposed his {\em instant form}, he~eliminated all time-like excitations, but he forgot to mention
his c-number time-energy uncertainty relation he formulated in
his earlier paper of 1927~\cite{dir27}.

Dirac's wife was Eugene Wigner's younger sister.  Dirac thus had
many occasions to meet his brother-in-law.  Dirac sometimes
quoted Wigner's paper on the inhomogeneous Lorentz group~\cite{wig39},
but without making any serious references, in spite of the fact
that Wigner's $O(3)$-like little group is the same as his own
instant form mentioned in his 1949 paper~\cite{dir49}.

Paul A. M. Dirac is an important person in the history of physics.
It is thus important to examine what conclusions we can draw if
we integrate all of those papers by closing up their soft spots.

In Section~\ref{efforts}, we list four important papers Dirac published
from 1927 to 1963~\cite{dir27,dir45,dir49,dir63}.  We then point out
his original ideas contained therein.

In 1971~\cite{fkr71}, Feynman, Kislinger, and Ravndal published
a paper saying that although quantum field theory works for scattering
problems with running waves, harmonic oscillators may be useful for
studying bound states in the relativistic world. They then formulated
a Lorentz-invariant differential equation separable into
a Klein--Gordon equation for a {\em free} hadron, and a Lorentz-invariant
oscillator equation for the bound state of the {\em quarks}.  However,
their solution of the oscillator equation is not normalizable and
is physically meaningless. In Section~\ref{fkr} we discuss their paper.

In Section~\ref{cobound}, we construct normalizable
harmonic oscillator wave functions.  These wave functions are not
Lorentz-invariant, because the shape of the wave function changes as
it is Lorentz-boosted.  However,
the wave function is Lorentz-covariant under Lorentz transformations.
It is shown further that these Lorentz-covariant
wave functions constitute a representation of Wigner's $O(3)$-like
little group for massive particles~\cite{wig39}.

In Section~\ref{quarkmo}, the covariant oscillator wave functions
are applied to hadrons moving with relativistic speed.  It is noted
that the wave function becomes squeezed along Dirac's light-cone
system~\cite{dir49}.  It~is shown that this squeeze property is
responsible for the dipole cut-off behavior of the proton form
factor~\cite{frazer60}.  It is shown further that Gell-Mann's
quark model~\cite{gell64} and Feynman's parton
model~\cite{fey69a,bj69} are two limiting cases of one
Lorentz-covariant entity, as in the case of Einstein's
$E = mc^2$~\cite{kn77par,kim89}.

In Section~\ref{twopho}, it is shown that the two-variable covariant
harmonic oscillator wave function can serve as the formula for
two-photon coherent states commonly called squeezed states of
light~\cite{yuen76}.  It is then noted that Dirac, in
his 1963 paper~\cite{dir63}, constructed the Lie algebra
of two-photon states with ten generators.
Dirac noted further that these generators satisfy the Lie algebra
of the $O(3,\,2)$ de~Sitter~group.

The $O(3,\,2)$ de Sitter group is a Lorentz group applicable to three
space-like dimensions and two time-like dimensions.   There are ten
generators for this group.  If we restrict ourselves to one of the
time-variables, it becomes the familiar Lorentz group with three
rotation and three boost generators.  These six generators lead to
the Lorentz group familiar to us.  The remaining four generators
are for three Lorentz boosts with respect to the second time variable
and one rotation generator between the two time variables.

In Section~\ref{contrac}, we contract these last four generators into four
space-time translation generators.  Thus, it is possible to transform
Dirac's $O(3,\,2)$ group to the inhomogeneous Lorentz group~\mbox{\cite{bkn19sym,bkn19qr}}.
In~this way, we show that quantum mechanics and special relativity
share the same symmetry ground.  Based on Dirac's four papers listed
in this section, we venture to say that this was Dirac's ultimate~purpose.

\section{Dirac's Efforts to Make Quantum Mechanics
Lorentz-Covariant}\label{efforts}

Paul A. M. Dirac made it his lifelong effort to formulate quantum
mechanics so that it would be consistent with special relativity.
In this section, we review four of his major papers on this subject.
In~each of these papers, Dirac points out fundamental difficulties
in this problem.

Dirac noted, in 1927~\cite{dir27}, that the emission of photons from
atoms is a manifestation of the uncertainty relation
between the time and energy variables. He also noted that,
unlike the uncertainty relation of Heisenberg
which allows quantum excitations, the time or energy axis has no
excitations along it. Hence, when attempting to combine these two
uncertainty relations in the
Lorentz-covariant world, there remains the serious difficulty that the
space and time variables are linearly mixed.

Subsequently in 1945~\cite{dir45}, Dirac, using the four-dimensional
harmonic oscillator attempted to construct, using the oscillator wave
functions, a representation of the Lorentz group. When he did this,
however, the wave functions which resulted did not appear to be
Lorentz-covariant.

Using the ten generators of the inhomogeneous Lorentz group
Dirac in 1949~\cite{dir49}, constructed from them
three forms of relativistic dynamics.  However, after imposing subsidiary conditions
necessitated by the existing form of quantum mechanics, he found
inconsistencies in all three of the forms he considered.

In 1963~\cite{dir63}, Dirac constructed a representation of the
$O(3,\,2)$ de Sitter group. To accomplish this, he used two coupled
harmonic oscillators in the form of step-up and step-down
operators. Thus Dirac constructed a beautiful algebra. He did not,
however, attempt to exploit the physical contents of his algebra.

In spite of the shortcomings mentioned above, it is indeed remarkable
that Dirac worked so tirelessly on this important subject.  We are
interested in combining all of his works to achieve his goal of making
quantum mechanics consistent with special relativity.  Let us review
the contents of these papers in detail, by transforming Dirac's
formulae into geometrical figures.

\subsection{Dirac's C-Number Time-Energy Uncertainty Relation}\label{quantu}

It was Wigner who in 1972~\cite{wig72} drew attention to the fact that the
time-energy uncertainty relation, known from the
transition time and line broadening in atomic spectroscopy,
existed before 1927. This~occurred even
before the uncertainty principle that Heisenberg formulated in 1927.
Also in 1927~\cite{dir27}, Dirac studied the uncertainty
relation which was applicable to the time and energy variables. When
the uncertainty relation was formulated by Heisenberg, Dirac
considered the possibility of whether a Lorentz-covariant uncertainty
relation could be formed out of the two uncertainty
relations~\cite{dir27}.

Dirac then noted that the time variable is a c-number and thus there are
no excitations along the time-like direction. However, there are
excitations along the space-like longitudinal direction starting from
the position-momentum uncertainty. Since the
space and time coordinates are mixed up for moving
observers, Dirac wondered how this
space-time asymmetry could be made consistent with Lorentz
covariance. This was indeed a major difficulty.

Dirac, however, never addressed, even in his later papers, the
separation in time variable or the time interval. On the other hand,
the Bohr radius, which measures the distance between the proton and
electron is an example of Heisenberg's uncertainty relation,
applicable to space separation variables.

In his 1949 paper~\cite{dir49} Dirac discusses his {\em instant form}
of relativistic dynamics. Thus Dirac came back to this question of the
space-time asymmetry, in his 1949 paper.  There he addresses
indirectly the possibility of freezing three of the six parameters of
the Lorentz group, and hence only working with the remaining three
parameters.  Wigner, in this 1939 paper~\cite{wig39,knp86} already
presented this idea.  He had observed in that paper that his little
groups with three independent parameters dictated the internal
space-time symmetries of particles.

\subsection{Dirac's Four-Dimensional Oscillators}

Since the language of special relativity is the Lorentz group, and
harmonic oscillators provide a start for the present form of quantum
mechanics, Dirac, during the second World War, considered~the
possibility of using harmonic oscillator wave functions to construct
representations of the Lorentz group~\cite{dir45}.  He considered
that, by constructing representations of the Lorentz group using
harmonic oscillators, he might be able to make quantum mechanics
Lorentz-covariant.

Thus in his 1945 paper~\cite{dir45}, Dirac considers the Gaussian form
\begin{equation}\label{ground4}
\exp\left(- \frac{1}{2}\left[x^2 + y^2 + z^2 + t^2\right]\right) .
\end{equation}
The $x$ and $y$ variables can be dropped from this expression, as we are
considering a Lorentz boost only along the $z$ direction. Therefore we
can write the above equation as:
\begin{equation}\label{ground}
\exp\left(- \frac{1}{2}\left[z^2 + t^2\right]\right) .
\end{equation}
Since $\left(z^2 - t^2\right)$ is an invariant quantity, the above
expression may seem strange for those who believe in Lorentz
invariance.

In his 1927 paper~\cite{dir27} Dirac proposed the time-energy
uncertainty relation, but observed that. because time is a c-number,
there are no excitations along the time axis. Hence the above
expression is consistent with this earlier paper.

\begin{figure}
\centerline{\includegraphics[width=10cm]{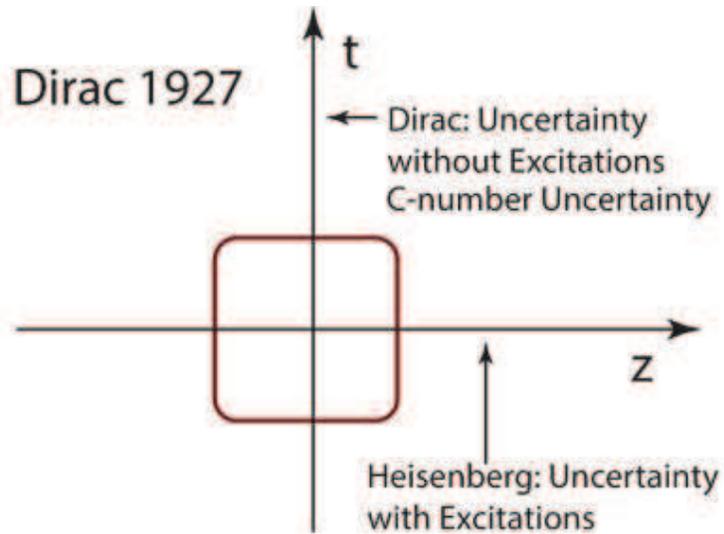}}
\vspace{5mm}
\caption{Quantum mechanics represented in terms of space-time.
As can be seen, there are no excitations along the time-like
direction, but quantum excitations along the space-like longitudinal
direction are allowed.}\label{diracf27}
\end{figure}

\begin{figure}
\centerline{\includegraphics[width=12cm]{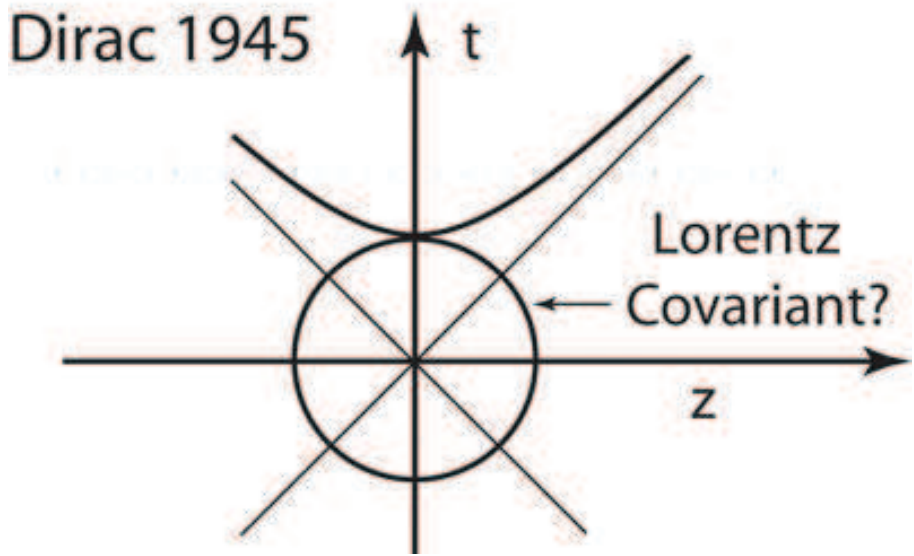}}
\caption{Dirac's four-dimensional oscillators localized in a closed
space-time region.  This is not a Lorentz-invariant concept.  How about
Lorentz covariance?}\label{diracf45}
\end{figure}

If we look carefully at Figure~\ref{diracf27}, we see that this figure
is a pictorial illustration of Dirac's Equation~(\ref{ground}).
There is localization in both space and time coordinates.  Dirac's
fundamental question, illustrated in Figure~\ref{diracf45}, would
then be how to make this figure covariant. Dirac stops there.
However,~this~is not the end of his story.

\subsection{Dirac's Light-Cone Coordinate System}\label{lightcone}

\begin{figure}
\centerline{\includegraphics[scale=0.82]{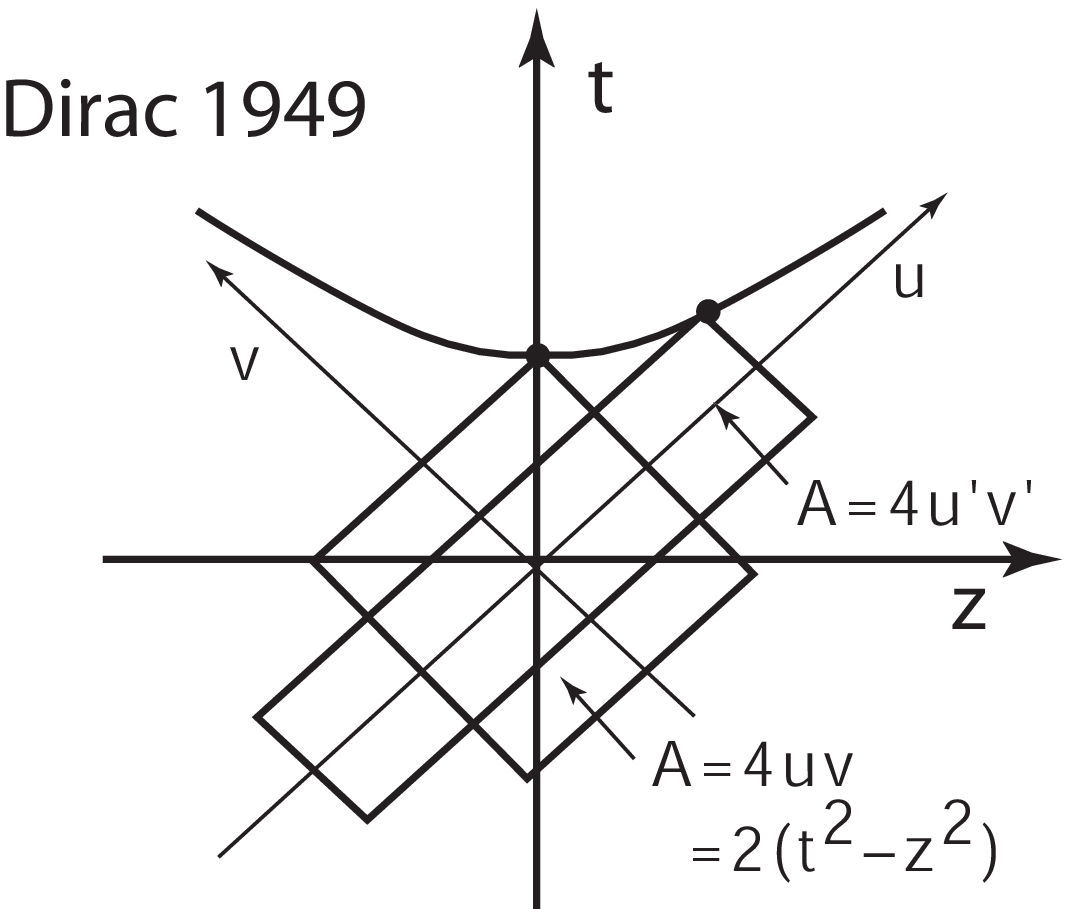}}
\vspace{2mm}
\caption{The light-cone coordinate system pictured with a Lorentz boost.
Not only does the boost squeeze the square into a rectangle, but
it traces a point along the hyperbola.}\label{diracf49}
\end{figure}

The Reviews of Modern Physics, in 1949, celebrated Einstein's
70th birthday by publishing a special issue.
In this issue was included Dirac's paper
entitled {\em Forms of Relativistic Dynamics}~\cite{dir49}.
Here Dirac introduced his light-cone coordinate system. In this
system a Lorentz boost is seen to be a squeeze transformation, where one
axis expands while the other contracts in such a way that their product
remains invariant as shown in Figure~\ref{diracf49}.

When boosted along the $z$ direction, the system is transformed
into the following form:
\begin{equation}\label{boostm}
\pmatrix{z' \cr t'  }  =
 \pmatrix{\cosh(\eta) & \sinh(\eta) \cr
  \sinh(\eta) & \cosh(\eta)}
\pmatrix{z \cr t } .
\end{equation}
Dirac defined his light-cone variables as~\cite{dir49}
\begin{equation}\label{lc11}
z_{+} = \frac{z + t}{\sqrt{2}} , \qquad z_{-} = \frac{z - t}{\sqrt{2}} .
\end{equation}
Then the form of the boost transformation of Equation~(\ref{boostm}) becomes
\begin{equation}\label{lorensq}
\pmatrix{ z_{+}' \cr z_{-}'} =
\pmatrix{ e^{\eta} & 0 \cr 0 & e^{-\eta } }
\pmatrix{ z_{+} \cr z_{-} } .
\end{equation}
It is then apparent that $u$ variable becomes expanded, but the $v$ variable becomes
contracted. We~illustrate this in Figure~\ref{diracf49}.  The product
then becomes:
\begin{equation}
z_{+}z_{-} = \frac{1}{2}(z + t)(z - t) = \frac{1}{2}\left(z^2 - t^2\right)
\end{equation}
which remains invariant.  The Lorentz boost is therefore, in Dirac's picture, a
squeeze transformation.

Dirac also introduced his 1949 paper his {\em instant form} of relativistic
quantum mechanics. This has the condition
\begin{equation}
        x_{0} \approx 0 .
\end{equation}
What did his approximate equality mean?  In this paper, we interpret
the nature of the time-energy uncertainty relation in terms of his c-number.
Furthermore, it could mean that it is, for the massive particle
is the three-dimensional rotation group,
Wigner's little group, without the time-like direction.

Additionally, Dirac stated that constructing a representation of
the inhomogeneous Lorentz group, was necessary to
construct a relativistic quantum mechanics. The
inhomogeneous Lorentz group has ten generators, four space-time translation generators,
three rotation generators, and three boost generators,
which satisfy a closed set of commutation relations.

It is now clear that Dirac was interested in using harmonic
oscillators to construct a representation of the inhomogeneous Lorentz
group. In 1979, together with another author, the present
authors published a paper on this oscillator-based
representation~\cite{kno79jmp}.  We regret that we did not mention
there Dirac's earlier efforts along this line.

\section{Scattering and Bound States}\label{fkr}

From the three papers written by Dirac~\cite{dir27,dir45,dir49},
let us find out what he really had in mind.  In~physical systems,
there are scattering and bound states.  Throughout his papers,
Dirac did not say he was mainly interested in localized bound
systems.  Let us clarify this issue using the formalism of Feynman,
Kislinger, and Ravndal~\cite{fkr71}.

We are quite familiar with the Klein--Gordon equation for a free particle
in the Lorentz-covariant world.  We shall use the four-vector notations
\begin{equation}
x_{\mu} = (x, y, z, t), \quad\mbox{and}\quad
                           x_{\mu}^2 =  x^2  + y^2 + z^2 - t^2 .
\end{equation}
Then the Klein--Gordon equation becomes
\begin{equation}
  \left(-\left[\frac{\partial}{\partial x_\mu}\right]^2
+ m^2 \right)\phi(x) = 0.
\end{equation}
The solution of this equation takes the familiar form
\begin{equation}
 \exp\left[\pm i \left(\ p_1 x + p_2 y + p_3 z \pm E t \right)\right] .
\end{equation}

In 1971, Feynman et al.~\cite{fkr71} considered two particles
a and b bound together by a harmonic oscillator potential, and wrote
down the equation
\begin{equation}\label{301}
\left\{-\left[\frac{\partial}{\partial x_{a\mu}}\right]^2 -
 \left[\frac{\partial}{\partial x_{b \mu}}\right]^2 +
\left(x_{a\mu} - x_{b\mu}\right)^2  + m_a^2 + m_b^2\right\}
\phi\left(x_{a\mu}, x_{b\mu}\right) = 0 .
\end{equation}

The bound state of these two particles is one {\em hadron}.  The
constituent particles are called {\em quarks}.
We can then define the four-coordinate vector of the hadron as
\begin{equation}
X = \frac{1}{2}\left(x_a + x_b\right),
\end{equation}
and the space-time separation four-vector between the quarks as
\begin{equation}
\quad x =\frac{1}{2\sqrt{2}} \left(x_a - x_b\right).
\end{equation}
Then  Equation~(\ref{301}) becomes
\begin{equation}
\left\{ -\left[\frac{\partial}{\partial X_{\mu}}\right]^2 + m_0^2
  + \left(-\left[\frac{\partial}{\partial x_{\mu}}\right]^2
             + x_{\mu}^2\right) \right\}\phi(X,x) = 0 .
\end{equation}
This differential equation can then be separated into
\begin{equation}\label{105}
\left(- \left[\frac{\partial}{\partial X_\mu}\right]^2
  + m_0^2 \right)\phi(X,x)
  = -\left(-\left[\frac{\partial}{\partial x_\mu}\right]^2
  + x_{\mu}^2\right) \phi(X,x),
\end{equation}
with
\begin{equation}\label{110}
\phi(X,x) = f(X)\psi(x) ,
\end{equation}
where $f(X)$ and $\psi(x)$ satisfy their own equations:
\begin{equation}
\left(-\left[\frac{\partial}{\partial X_\mu}\right]^2 +
 m_{a}^2 + m_{b}^2 + \lambda \right) f(X) = 0
\end{equation}
and
\begin{equation}\label{112}
 \frac{1}{2}\left(-\left[\frac{\partial}{\partial x_{\mu}}\right]^2
  + x_{\mu}^2\right)\psi(x) = \lambda \psi(x) .
\end{equation}

Here, the wave function then takes the  form
\begin{equation}\label{wf11}
\phi(X,x) = \psi(x)
 \exp\left[\pm i\left( P_x X + P_y Y + P_z Z \pm E T\right)\right] ,
\end{equation}
where $P_x, P_y, P_z$ are for the hadronic momentum, and
\begin{equation}
E^2 = P_x^2   + P_y^2  + P_z^2 + M^2,
\quad\mbox{with}\quad M^2 = m_{a}^2 + m_{b}^2 + \lambda .
\end{equation}
Here the hadronic mass $M$ is determined by the parameter $\lambda$,
which is the eigenvalue of the differential equation for $\psi(x)$
given in Equation~(\ref{112}).

\begin{figure}
\centerline{\includegraphics[width=9cm]{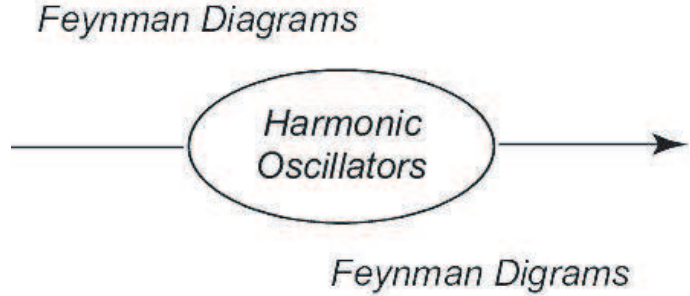}}
\vspace{5mm}
\caption{Feynman, in an effort to combine quantum mechanics with special
relativity, gave us this roadmap.  Feynman's diagrams provide, in
Einstein's world, a satisfactory resolution for scattering states.
Thus they work for running waves.
Feynman suggested that harmonic oscillators should be used as a first
step for representing standing waves trapped inside an extended
hadron.}\label{dff33}
\end{figure}

Considering Feynman diagrams based on the S-matrix formalism, quantum
field theory has been quite successful. It is, however, only useful
for physical processes where, after interaction, one set of free
particles becomes another set of free
particles.  The questions of localized probability distributions and their covariance
under Lorentz transformations is not addressed by quantum field
theory.  In order to tackle this
problem and address these questions, Feynman et al. suggested
harmonic oscillators~\cite{fkr71}.  In Figure~\ref{dff33}, we illustrate
this idea.

However, for their wave function $\psi(x)$, Feynman et al.
uses a Lorentz-invariant exponential form
\begin{equation}
\exp\left( -\frac{1}{2}\left[x^2 + y^2 + z^2 - t^2\right]\right).
\end{equation}
This wave function increases as $t$ becomes large.  This is not
an acceptable wave function.  They~overlooked the normalizable
exponential form given by Dirac in Equation~(\ref{ground4}). They
also overlooked the form in the paper of Fujimura et al.~\cite{fuji70}
which was quoted in their own paper.

Thus, we are interested in fixing this problem and thus constructing
Lorentz-covariant oscillator wave functions satisfying both the
rules of quantum mechanics and the rules of special relativity.

\section{Lorentz-Covariant Picture of Quantum Bound States}\label{cobound}

In 1939, Wigner considered internal space-time symmetries of particles
in the Lorentz-covariant world~\cite{wig39}.  For this purpose, he
considered the subgroups of the Lorentz group for a given four-momentum
of the particle.  For the massive particle, the internal space-time
symmetry is defined for when the particle is at rest, and its symmetry
is dictated by the three-dimensional rotation group, which allows us to
define the particle spin as a dynamical variable, as indicated in
Table~\ref{further}.

Let us go to the wave function of Equation~(\ref{wf11}). This wave function
$\psi(x)$ is for the internal coordinates of the hadron, and the
exponential form defines the hadron momentum.  Thus, the symmetry
of Wigner's little group is applicable to the wave function $\psi(x)$.

This situation is like the case of the Dirac equation for a free particle.
Its solution is a plane wave times the four-component Dirac spinor
describing the spin orientation and its Lorentz covariance.  This~separation
of variables for the present case of relativistic extended
particles is illustrated in Figure~\ref{dff33}.  With this understanding
let us write the Lorentz-invariant differential equation as
\begin{equation}\label{425}
\frac{1}{2}\left[-\frac{\partial^2}{\partial x^2} -
\frac{\partial^2}{\partial y^2} - \frac{\partial^2}{\partial z^2}
   + \frac{\partial^2}{\partial t^2}
   + \left(x^2 + y^2 + z^2 - t^2\right) \right]\psi(x,y,z,t)
   = \lambda \psi(x,y,z,t) .
\end{equation}

Here, the variables $x,\, y,\, z$ are for the spatial separation
between the quarks, like the Bohr radius in the hydrogen atom.
The time variable $t$ is the {\em time separation} between the
quarks.  This variable is very strange, because it does not exist
in the present forms of quantum mechanics and quantum field theory.
Paul A. M. Dirac did not mention this time separation in any of
his papers quoted here.  Yet,~it plays the major
role in the Lorentz-covariant world, because the spatial separation
(like the Bohr radius) picks up a time-like component when
the system is Lorentz-boosted~\cite{kn06aip}.

In his 1927 paper~\cite{dir27}, Dirac mentioned the c-number
time-energy uncertainty relation, and he used $t \approx 0$ in his
1949 paper for his instant form of relativistic dynamics.  When
he wrote down a Gaussian function with $\left(x^2 + y^2 + z^2 + t^2\right)$
as the exponent, he should have meant $x,\, y$ and $z$ are for the
space separation and $t$ for the time separation,  since otherwise
the system becomes zero in the remote future and remote past.

With this understanding, we are dealing here with the solution of
the differential equation of the~form
\begin{equation}\label{401}
\psi(x) = f(x, y, z) \exp\left(\frac{-t^2}{2}\right),
\end{equation}
where $f(x, y, z)$ satisfy the oscillator differential equation
\begin{equation}\label{135}
\frac{1}{2}\left(-\frac{\partial^2}{\partial x^2} - \frac{\partial^2}{\partial y^2}
 -\frac{\partial^2}{\partial z^2}
 + x^2 + y^2 + z^2 \right)f(x,y,z) = \left(\lambda - \frac{1}{2}\right)f(x,y,z) .
\end{equation}

The form of Equation~(\ref{401}) tells us there are no time-like excitations.
This equation is the Schr\"odinger equation for the three-dimensional
harmonic oscillator and its solutions are well known.

If we use the three-dimensional spherical coordinate system, the
solution will give the spin or internal angular momentum and the
orientation of the bound-state hadron~\cite{kno79jmp}.  This spherical
form is for the $O(3)$ symmetry of Wigner's little group for
massive particles.  The Lorentz-invariant Casimir operators are
given in Ref.~\cite{kno79jmp}.

If we are interested in Lorentz-boosting the wave function, we
note that the original wave equation of Equation~(\ref{425}) is separable
in all four variables.  If the Lorentz boost is made along the $z$
direction, the wave functions along the $x$ and $y$ directions
remain invariant, and thus can be separated.  We can study only the
longitudinal and time-like components.  Thus the differential equation of
Equation~(\ref{425}) is reduced to
\begin{equation}\label{435}
\frac{1}{2}\left[-\frac{\partial^2}{\partial z^2} + \frac{\partial^2}{\partial t^2}
  + \left( z^2 - t^2\right) \right]\psi(z,t)  = \lambda \psi(z,t) .
\end{equation}

The solution of this differential equation takes the form

\begin{equation}
   \psi(z,t) = H_{n}(z)\exp\left(-\left[\frac{z^2 + t^2}{2}\right]\right),
\end{equation}
where $H_{n}(z)$ is the Hermite polynomial. There are no excitations
in $t$, therefore it is restricted to the ground state.  For
simplicity, we ignore the normalization constant.

If this wave function is Lorentz-boosted along the $z$ direction,
the $z$ and $t$ variables in this expression should be replaced
according to
\begin{equation}\label{440}
  z \rightarrow (\cosh\eta)z - (\sinh\eta)t, \quad\mbox{and}\quad
  t \rightarrow (\cosh\eta)t - (\sinh\eta)z.
\end{equation}
According to the light-cone coordinate system introduced by Dirac
in 1949~\cite{dir45}, this transformation can be written as
\begin{equation}
(z + t) \rightarrow e^{-\eta}(z + t), \quad\mbox{and}\quad
(z - t) \rightarrow e^{\eta} (z - t).
\end{equation}
Thus the Lorentz-boosted wave function  becomes
\begin{equation}
\psi_{\eta}(z,t) = H_{n}\left(\frac{1}{\sqrt{2}}
 \left[ e^{-\eta} (z + t) + e^{\eta}(z - t)\right]\right)
 \exp\left(-\frac{1}{4}\left[e^{-2\eta }(z + t)^{2} +
   e^{2\eta}(z - t)^{2}\right]\right) ,
\end{equation}
without the normalization constant.

It is possible to write this wave function using the one-dimensional
normalized oscillator functions $\phi_{n}(z)$ and $\phi_{k}(t)$ as
\begin{equation}
 \psi_{\eta}(z,t) = \sum_{nk} A_{nk} \phi_{n}(z)\phi_{k}(t) .
\end{equation}
This problem has been extensively discussed in the
literature~\cite{knp86,ruiz74,kno79ajp,rotbart81}.

\begin{figure}
\centerline{\includegraphics[width=12cm]{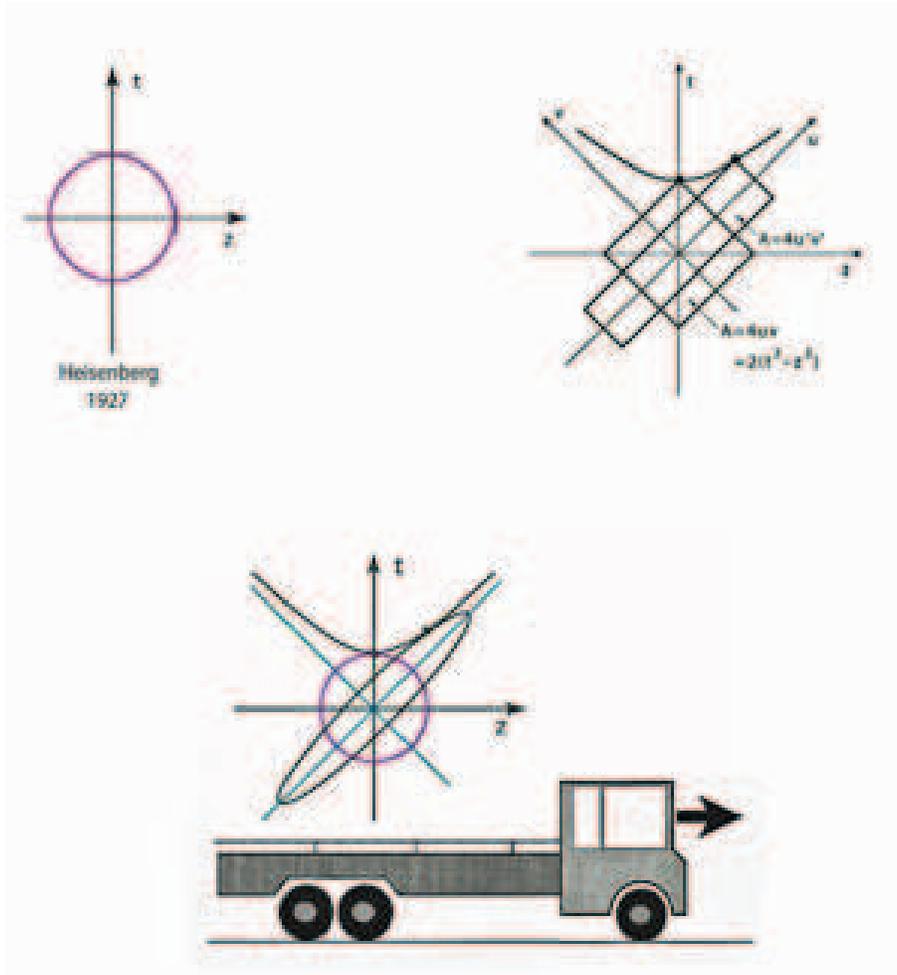}}
\vspace{5mm}
\caption{Lorentz covariance of the internal the wave function
$\psi_{\eta}(z,t)$ of Equation~(\ref{wf44}).
Synthesizing~Figures~\ref{diracf45} and \ref{diracf49}, we obtain
a squeezed circle as shown in this figure.  This figure thus
integrates Dirac's three
papers~\cite{dir27,dir45,dir49}.}\label{truck}
\end{figure}

The most interesting case is the expansion of the ground state.
The normalized wave function is~then
\begin{equation}\label{wf44}
\psi_{\eta }(z,t) = \left(\frac{1}{\pi}\right)^{1/2}
 \exp\left(-\frac{1}{4}\left[e^{-2\eta }(z + t)^{2} +
  e^{2\eta}(z - t)^{2}\right]\right) .
\end{equation}
The Lorentz-boost property of this form is illustrated in Figure~\ref{truck}.
The expansion in the harmonic oscillator wave functions takes the form
\begin{equation}
 \psi_{\eta}(z,t) = \frac{1}{\cosh\eta} \sum_{n} (\tanh\eta)^{n}
       \phi_{n}(z)\phi_{n}(t) .
\end{equation}
This expression is the key formula for two-photon coherent states
or squeezed states of light~\cite{yuen76}.  We~shall return
to this two-photon problem in Section~\ref{twopho}.

As for the wave function $\psi(x)$, this is the localized wave function
Dirac was considering in his papers of 1927 and 1945.  If
Figures~\ref{diracf45} and \ref{diracf49} are combined, we end
up with an ellipse as a squeezed circle as shown in Figure~\ref{truck}.

Indeed, one of the most controversial
issues in high-energy physics is explained by this squeezed circle.
The bound-state quantum mechanics of protons, which are known to be bound states
of quarks, are often assumed to be the
same as that of the hydrogen atom.  Then how would the proton
look to an observer on a train becomes the question. According to
Feynman~\cite{fey69a,bj69}, when the speed of the train
becomes close to that of light, the proton appears like a collection
of partons. However,~the~properties of Feynman's
partons are quite different from the properties  of the quarks.
This issue shall be discussed in more detail in Section~\ref{quarkmo}.

\section{Lorentz-Covariant Quark Model}\label{quarkmo}

Early successes in the quark model include the calculation of the
ratio of the neutron and proton magnetic moments~\cite{beg64}, and
the hadronic mass spectra~\cite{fkr71,owg67}.  These are based on
hadrons at rest.  We are interested in this paper how the hadrons
in the quark model appear to observers in different Lorentz frames.

These days, modern particle accelerators routinely produce protons
moving with speeds very close to that of light.  Thus, the question
is therefore whether the covariant wave function developed in
Section~\ref{cobound} can explain the observed phenomena associated
with those protons moving with relativistic speed.

The idea that the proton or neutron has a space-time extension had
been developed long before Gell-Mann's proposal for the quark
model~\cite{gell64}.  Yukawa~\cite{yuka53} developed this idea as
early as 1953, and~his idea was followed up by Markov~\cite{markov56},
and by Ginzburg and Man'ko~\cite{ginz65}.

Since Einstein formulated his special relativity for point particles,
it has been and still is a challenge to formulate a theory for
particles with space-time extensions.  The most naive idea would
be to study rigid spherical objects, and there were many papers
on this subjects.  But we do not know where that story stands
these days.  We can however replace these extended rigid bodies
by extended wave packets or standing waves, thus by localized
probability entities.  Then what are the constituents within
those localized waves?  The quark model gives the natural answer
to this question.

Hofstadter and McAllister~\cite{hofsta55}, by using
electron-proton scattering to measure the charge distribution
inside the proton, made the first experimental discovery of the
non-zero size of the proton.
If the proton were a point particle, the
scattering amplitude would just be a Rutherford formula.  However,
Hofstadter and MacAllister found a tangible departure from this
formula which can only be explained by a spread-out charge
distribution inside the proton.

In this section, we are interested in how well the bound-state
picture developed in Section~\ref{cobound} works in explaining
relativistic phenomena of those protons.  In Section~\ref{formfac}
we study in detail how the Lorentz squeeze discussed in
Section~\ref{cobound} can explain the behavior of electron-proton
scattering as the momentum transfer becomes relativistic.

Second, we note that the proton is regarded as a bound state
of the quarks sharing the same bound-state quantum mechanics
with the hydrogen atom.  However, it appears as a collection
of Feynman's partons. Thus, it is a great challenge to see
whether one Lorentz-covariant formula can explain both the
static and light-like protons.  We shall discuss this issue
in Section~\ref{fparton}

One hundred years ago, Einstein and Bohr met occasionally to
discuss physics.  Bohr was interested in how the election orbit
looks and Einstein was worrying about how things look to moving
observers.  Did they ever talk about how the hydrogen atom
appears to moving observers?  In~Section~\ref{histo}, we
briefly discuss this issue.

It is possible to conclude from the previous discussion that the
Lorentz boost might increase the uncertainty. To deal with this,
in Section~\ref{uncert}, we address the issue of the uncertainty
relation when the oscillator wave functions are Lorentz-boosted.

\subsection{Proton Form Factor}\label{formfac}

Using the Born approximation for non-relativistic scattering,
we see what effect the charge distribution has on the
scattering amplitude.
When electrons are scattered from a fixed charge
distribution with a  density of $e\rho(r)$, the scattering amplitude
becomes:
\begin{equation}\label{501}
f(\theta) = - \left(\frac{e^{2}m}{2\pi}\right)
 \int d^3x d^3x' \frac{\rho(r')}{R} \exp{(-i {\bf Q}\cdot {\bf x})} .
\end{equation}
Here we use
$r = |{\bf x}|,  R =|{\bf r} - {\bf r'} |, $ and ${\bf Q} = {\bf K_f}
- {\bf K_i},$  which is the momentum transfer.  We can reduce this amplitude
to:
\begin{equation}\label{502}
f(\theta) = \frac{2me^2}{Q^2} F(Q^2) .
\end{equation}
The density function's Fourier transform is given by $F(Q^2)$ which can
be written as:
\begin{equation}\label{503}
F\left(Q^2\right) = \int d^3x \rho(r) \exp{(-i {\bf Q}\cdot {\bf x})} .
\end{equation}
This above quantity is called the form factor and it describes the charge
distribution in terms of the momentum transfer which can be
normalized by:
\begin{equation}\label{504}
\int \rho(r) d^3x = 1.
\end{equation}
Therefore, from Equation~(\ref{503}), F(0) = 1.
the scattering amplitude of Equation~(\ref{501})
becomes the Rutherford formula for Coulomb scattering
if the density function, corresponding to a point charge, is a delta
function, $F\left(Q^2\right) = 1$, for all values of $Q^2$. Increasing
values of $Q^2$,  which are deviations
from Rutherford scattering, give a measure
of the charge distribution. Hofstadter's experiment, which~scattered
electrons from a proton target, found this precisely~\cite{hofsta55}.

\begin{figure}
\centerline{\includegraphics[width=8cm]{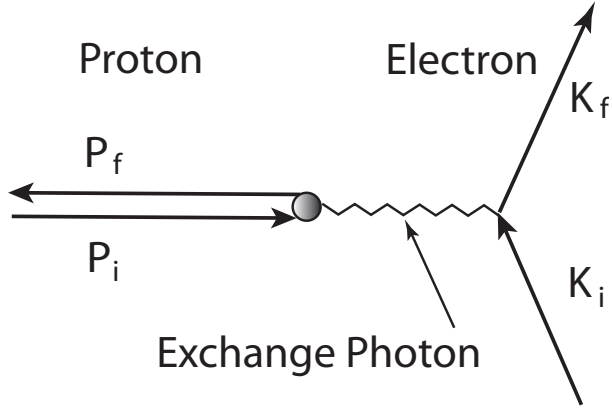}}
\vspace{5mm}
\caption{Electron-proton scattering in the Breit frame.  The
outgoing momentum of the proton is opposite in sign but equal in magnitude to
that of the incoming proton. }\label{breit}
\end{figure}

When the energy of the incoming electron becomes higher, it is necessary to take
into account the recoil effect of target proton. It then requires that
the problem be formulated in the Lorentz-covariant framework.
It is generally agreed that quantum electrodynamics can describe
electrons and their electromagnetic interaction by using Feynman
diagrams for practical calculations.
When using perturbation, a power series of the fine structure constant
$\alpha.$ is used to expand the scattering amplitude.
Therefore,
the lowest order in $\alpha,$ can, using the diagram given in
Figure~\ref{breit}, describe the scattering of an
electron by a proton.

Many textbooks on elementary particle physics~\cite{frazer66,itzykson80} give the
corresponding matrix element as the form:
\begin{equation}\label{505}
\bar{U}\left(P_f\right)\Gamma_{\mu}\left(P_f, P_i\right)
\left(\frac{1}{Q^2}\right)
\bar{U}\left(K_f\right)\gamma^{\mu}U\left(K_i\right) ,
\end{equation}
where the initial and final
four-momenta of the proton and electron, respectively, are given by
$P_i, P_f, K_i$ and $K_f$.
The Dirac spinor for the initial proton is $U\left(P_i\right)$, while
the (four-momentum transfer$)^2$, $Q^2$, is
\begin{equation}
      Q^2 = \left(P_f - P_i\right)^2 = \left(K_f - K_i\right)^2  .
\end{equation}

If the proton were a point particle like the electron, the
$\Gamma_{\mu}$ would be $\gamma_{\mu}$, but for a particle, like the
proton, with space-time
extension, it is
$\gamma_{\mu}F\left(Q^2\right)$.
The virtual photon being exchanged between the electron and the proton
produces the $\left(1/Q^2\right)$ factor in Equation~(\ref{505}).
For the particles involved in the
scattering process this quantity is positive for physical
values of the four-momenta in the metric we use.

From the definition of the form factor given in Equation~(\ref{503}),  we
can make  a relativistic calculation of the form factor.
The density function, which depends only on the target particle, is
proportional to $\psi(x)^{\dagger}\psi(x)$.  The
wave function for quarks inside the proton is $\psi(x)$. This expression is a
special case of the more general form
\begin{equation}
\rho(x) = \psi_f^{\dagger}(x)\psi_i(x) ,
\end{equation}
where the initial and final wave
function of the target atom is given by $\psi_i$  and  $\psi_f$.
The form factor of
Equation~(\ref{503}) can then
be written as
\begin{equation}
F\left(Q^2\right) =
  \left(\psi_{f}(x), e^{-i{\bf Q}\cdot{\bf r}} \psi_{i}(x)\right) .
\end{equation}

The required Lorentz generalization, starting from this expression,
can be made using the relativistic wave functions for hadrons.

By replacing each quantity in the expression of
Equation~(\ref{503}) by its relativistic counterpart we should be able to
see the details of the transition to relativistic physics.
If we go back to the
Lorentz frame in which the momenta of the incoming and outgoing
nucleons have equal magnitude but opposite signs, we obtain
\begin{equation}
{\bf p_i} + {\bf p_f} = 0 .
\end{equation}

This kinematical condition is illustrated in Figure~\ref{breit}.

We call the Lorentz frame in which the above condition holds the Breit
frame.  As illustrated
in Figure~\ref{breit}, there is no loss of generality if the proton
comes in along the $z$ direction before the collision and goes out along
the negative $z$ direction after the scattering process. The four
vector, $Q = (K_f - K_i) = (P_i - P_f)$, ín this frame, has no
time-like component. The Lorentz-invariant form,
$Q\cdot x$, can thus replace the exponential factor
${\bf Q}\cdot {\bf r}$.
The covariant harmonic oscillator wave functions discussed in this paper
can be used for the wave functions for the protons,
assuming that the nucleons are in the ground state.  Then
the integral in the evaluation of Equation~(\ref{503}),
which includes the time-like direction and is thus four-dimensional,
is the only
difference between the non-relativistic and relativistic cases.
The exponential factor, which does not depend on the time-separation variable,
therefore, is not affected by the integral in the time-separation variable.

We can now consider the integral:
\begin{equation}\label{506}
g\left(Q^2\right) = \int d^4x \psi^{\dagger}_{-\eta}(x)
  \psi_{\eta}(x) \exp{(-iQ\cdot x)} ,
\end{equation}
where $\tanh\eta$ is the velocity parameter ($ \tanh\eta = v/c $) for
the incoming proton, and
the wave function $\psi_{\eta}$, from Equation~(\ref{wf44}), takes the form:
\begin{equation}
\psi_{\eta}(x) = \frac{1}{\sqrt{\pi}}
\exp\left( -\frac{1}{4}\left[e^{-2\eta} (z + t)^2 +
                   e^{2\eta} (z - t)^2 \right] \right) .
\end{equation}

We can, now, after the above decomposition of the wave functions, perform
the integrations in the $x$ and $y$ variables trivially.  We can, after
dropping these trivial factors, write
the product of the two wave functions as
\begin{equation}\label{508}
\psi^{\dagger}_{-\eta}(x)\psi_{\eta}(x)
 = \frac{1}{\pi} \exp\left(- [\cosh(2\eta)]\left[t^2 + z^2\right]\right) .
\end{equation}
Now the $z$ and $t$ variables are separated.
Since the $t$ integral in
Equation~(\ref{506}), as the exponential
factor in Equation~(\ref{503}), does not depend on $t$, it can also be
trivially performed. Then integral
of Equation~(\ref{506}) can be written as
\begin{equation}\label{507}
g\left(Q^2 \right) = \frac{1}{\sqrt{\pi \cosh(2\eta)}}
  \int  e^{-2iPz} \exp\left(- \cosh(2\eta) z^2 \right) dz .
\end{equation}
Here the $z$ component of the momentum of the incoming proton is $P$.
The variable $Q^2$, which is the (momentum transfer$)^2$, now become
$4P^2$.
The hadronic material, which is distributed along the longitudinal direction,
has indeed became contracted~\cite{licht70}.

We note that $\tanh\eta$ can be written as
\begin{equation}
 (\tanh\eta)^{2} = \frac{Q^2}{Q^2 + 4M^2} ,
\end{equation}
where $M$ is the proton mass.  This equation tells us that $\beta = 0$ when
$Q^2 = 0$, while it becomes one as $Q^2$ becomes infinity.

The evaluation of the above integral for $g\left(Q^2\right)$ in
Equation~(\ref{507}) leads to
\begin{equation}
 g\left(Q^2\right) = \left(\frac{2M^2}{Q^2 + 2M^2}\right)
 \exp\left(\frac{-Q^2}{2(Q^2 + 2M^2)}\right) .
\end{equation}
For $Q^2 = 0,$ the above expression becomes 1.  It decreases as
\begin{equation}\label{509}
 g\left(Q^2\right) \sim  \frac{1}{Q^2}
\end{equation}
as $Q^2$ assumes large values.

So far the calculation has been performed for an oscillator bound
state of two quarks.  The proton, however, consists of three quarks.  As shown
in the paper of Feynman et al.~\cite{fkr71}, the problem becomes
a product of two oscillator modes. Thus, the three-quark system is
a straightforward generalization of the
above calculation. As a result, the form factor $F\left(Q^2\right)$ becomes;
\begin{equation}
 F\left(Q^2\right) = \left(\frac{2M^2}{Q^2 + 2M^2}\right)^2
     \exp\left(\frac{- Q^2}{Q^2 + 2M^2}\right) ,
\end{equation}
which is 1 at $Q^2 = 0$, and decreases as
\begin{equation}\label{510}
 F\left(Q^2\right) \sim \left[\frac{1}{Q^2}\right]^2
\end{equation}
as $Q^2$ assumes large values. This form factor function has the
required {\em dipole-cut-off} behavior, which has been observed in high-energy laboratories.
This calculation was carried first by Fujimura et al. in
1970~\cite{fuji70}.

Let us re-examine the above calculation.  If we replace $\beta$
by zero in Equation~(\ref{507}) and ignore the elliptic deformation of the
wave functions, $g\left(Q^2\right)$ will become
\begin{equation}
g\left(Q^2\right) = \exp\left(-Q^2/4\right) ,
\end{equation}
which will lead to an exponential cut-off of the form factor.  This
is not what we observe in laboratories.

In order to gain a deeper understanding of the above-mentioned
correlation, let us study the case using the momentum-energy wave
functions:
\begin{equation}
 \phi_{\eta}(q) = \left(\frac{1}{2\pi}\right)^2
 \int d^4x e^{-iq\cdot x} \psi_{\eta} (x) .
\end{equation}
If we ignore, as before, the transverse components, we can write
$g\left(Q^2\right)$ as~\cite{knp86}
\begin{equation}
 \int  dq_0 dq_z \phi^*_\eta \left(q_0, q_z - P\right)
 \phi_\eta \left(q_0, q_z + P \right) .
\end{equation}

\begin{figure}
\centerline{\includegraphics[width=8cm]{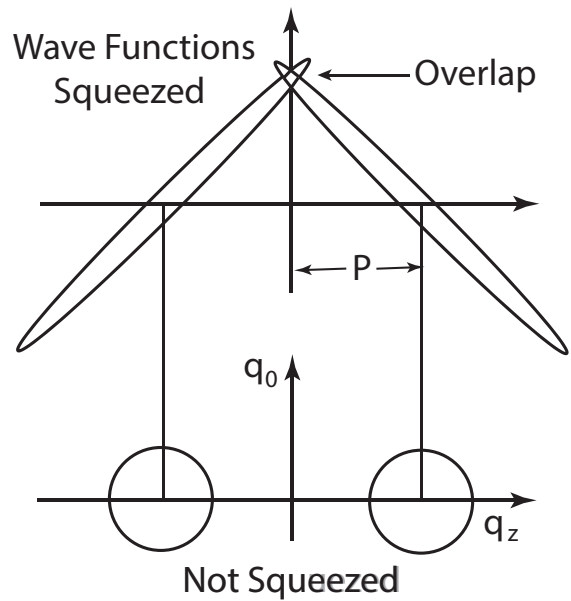}}
\vspace{5mm}
\caption{The momentum-energy wave functions are Lorentz squeezed in
the form factor calculation. The two wave functions become separated
as the momentum transfer increases.  However, in the relativistic
case, the wave functions maintain an overlapping region. In the
non-relativistic calculation, the wave functions
become completely separated.
The unacceptable behavior
of the form factor is caused by this lack of overlapping region.}\label{overlap}
\end{figure}

The above overlap integral has been sketched in Figure~\ref{overlap}.
The two wave functions overlap completely
in the $q_z \, q_0$ plane if $Q^2 = 0$ or $P = 0$.
The wave functions become
separated when P increases. Because of
the elliptic or squeeze deformation, seen in Figure~\ref{overlap}, they
maintain a small overlapping region.
In the non-relativistic case, there is no overlapping region, because
the deformation is not taken into account, as seen in
Figure~\ref{overlap}. The slower decrease in $Q^2$ is, therefore, more
precisely given in the relativistic
calculation than in the non-relativistic calculation.

Although our interest has been in the space-time behavior of the
hadronic wave function, it must be noted that quarks are spin-1/2
particles. This fact must be taken into consideration.  This spin
effect manifests itself prominently in the baryonic mass spectra. Here
we are concerned with the relativistic effects, therefore, it is
necessary to construct a relativistic spin wave function for the
quarks.  The result of this relativistic spin wave function
construction for the quark wave function should be a hadronic spin
wave function.  In the case of nucleons, the quark spins should be
combined in a manner to generate the form factor of Equation~(\ref{508}).

Naively we could use free Dirac spinors
for the quarks. However, it was shown by Lipes~\cite{lipes72} that
using free-particle Dirac spinors leads to a wrong behavior of the form
factor. Lipes' result, however,~does~not cause us any worry, as quarks in a
hadron are not free particles.
Thus we have to find
suitable mechanism in which quark spins, coupled to orbital
motion, are taken into account.  This is a difficult problem and is a
nontrivial research problem, and further study is needed along this direction~\cite{henriq75}.

In 1960, Frazer and Fulco calculated the form factor using the
technique of dispersion relations~\cite{frazer60}.  In so doing
they had to assume the existence of the so-called $\rho$ meson,
which was later found experimentally, and which subsequently played
a pivotal role in the development of the quark model.

Even these days, the form factor calculation occupies a very
important place in recent theoretical models, such as quantum
chromodynamics (QCD) lattice
theory~\cite{matevo05} and the Faddeev equation~\cite{roberts05}.
However, it is still noteworthy that Dirac's form of Lorentz-covariant
bound states leads to the essential dipole cut-off behavior of the
proton form factor.

\subsection{Feynman'S Parton Picture}\label{fparton}

As we did in Section~\ref{formfac}, we continue using the Gaussian form
for the wave function of the proton.  If the proton is at rest, the
$z$ and $t$ variables are separable, and the time-separation can be
ignored, as we do in non-relativistic quantum mechanics.  If the proton
moves with a relativistic speed, the wave function is squeezed as
described in Figure~\ref{truck}.  If the speed reaches that of light,
the wave function becomes concentrated along positive light cone with
$t = z$.  The question then is whether this property can explain the
parton picture of  Feynman when a proton moves with a speed close to
that of light.

It was Feynman who, in 1969, observed that a fast-moving proton can
be regarded as a collection of many {\em partons}. The properties of
these partons appear to be quite different from those of the
quarks~\cite{fey69a,bj69,knp86}.  For example, while
the number of quarks inside a static proton is three, the number
of partons appears to be infinite in a rapidly moving proton.
The following systematic observations were made by Feynman:

\begin{itemize}

\item[a.]  When protons move with
 velocity close to that of light, the parton picture is valid.

\item[b.] Partons behave as free independent particles when
  the interaction time between the quarks becomes dilated.

\item[c.] Partons have a widespread distribution of momentum
 as the proton moves quickly.

\item[d.] There seems to be an infinite number of partons or a number much larger
    than that of quarks.

\end{itemize}

\noindent  The question is whether the Lorentz-squeezed wave function
 produced in Figure~\ref{truck} can explain all of these peculiarities.

Each of the above phenomena appears as a paradox, when the proton is
believed to be a bound state of the quarks. This is especially true of
(b) and (c) together.  We can ask how a free particle can have a
wide-spread momentum distribution.

To resolve this paradox, we construct the
momentum-energy wave function corresponding to Equation~(\ref{wf44}).
We can construct two independent four-momentum variables~\cite{fkr71}
if the quarks have the four-momenta $p_{a}$ and $p_{b}$.
\begin{equation}
P = p_{a} + p_{b} , \qquad q = \sqrt{2}(p_{a} - p_{b}) .
\end{equation}
Since $P$ is the total four-momentum, it is the
four-momentum of the proton. The four-momentum separation
between the quarks is measured by $q$.  We can then write the
light-cone variables as
\begin{equation}\label{conju}
q_{+} = (q_{0} + q_{z})/\sqrt{2} , \quad
q_{-} = (q_{0} - q_{z})/\sqrt{2} .
\end{equation}
This results in the momentum-energy wave function
\begin{equation}\label{phi}
\phi_{\eta }(q_{z},q_{0}) = \left({1 \over \pi }\right)^{1/2}
\exp\left\{-{1\over 2}\left[e^{-2\eta}q_{+}^{2} +
e^{2\eta}q_{-}^{2}\right]\right\} .
\end{equation}

\begin{figure}
\centerline{\includegraphics[width=12cm]{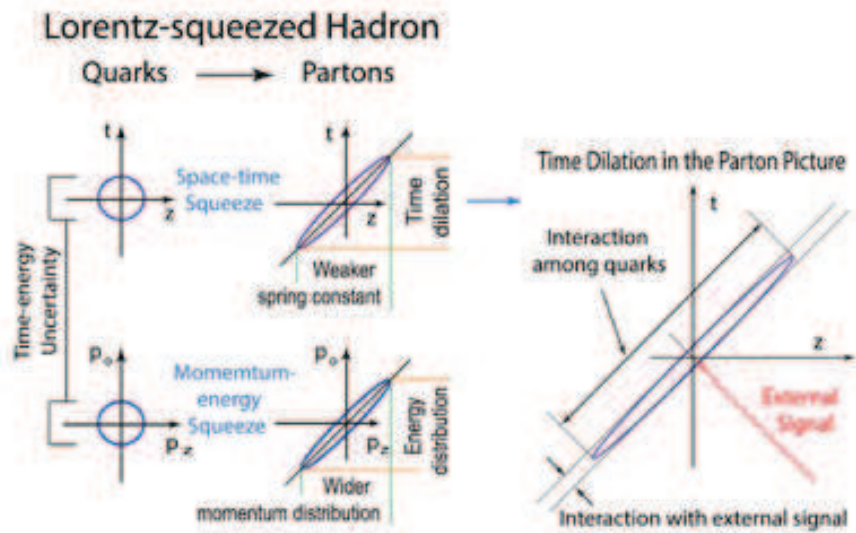}}
\caption{Lorentz-squeezed wave
functions in space-time and in momentum-energy variables. Both~wave
functions become concentrated along their respective positive
light-cone axes as the speed of the proton approaches that of light.
All the peculiarities of Feynman's
parton picture are presented in these light-cone concentrations.}\label{parton33}
\end{figure}

Since the harmonic oscillator is being used here, it is easily seen
that the above momentum-energy wave function has the identical
mathematical form to that of the space-time wave function of
Equation~(\ref{wf44}) and that these wave functions also have the same
Lorentz squeeze properties.  Though~discussed extensively in the
literature~\cite{knp86,kn77par,kim89}, these mathematical forms and
Lorentz squeeze properties are illustrated again in
Figure~\ref{parton33} of the present paper.

We can see from the figure, that both wave functions behave like those
for the static bound state of quarks when the proton is at rest with
$\eta = 0$. However, it can also be seen that as $\eta$ increases, the~wave functions become concentrated along their respective positive
light-cone axes as they become continuously squeezed.  If we look at the
$z$-axis projection of the space-time wave function, we see that, as
the proton speed approaches that of the speed of light, the width of
the quark distribution increases. Thus, to the observer in the
laboratory, the position of each quark appears widespread. Thus~the
quarks appear like free particles.

If we look at the momentum-energy wave function we see that it is just
like the space-time wave function. As the proton speed approaches that
of light, the longitudinal momentum distribution becomes
wide-spread. In non-relativistic quantum mechanics we expect that the
width of the momentum distribution is inversely proportional to that
of the position wave function.  This wide-spread longitudinal momentum
distribution thus contradicts our expectation from non-relativistic
quantum mechanics.  Our expectation is that free quarks must have a
sharply defined momenta, not a wide-spread momentum distribution.

However, as the proton is boosted, the space-time width and the
momentum-energy width increase in the same direction. This is because
of our Lorentz-squeezed space-time and momentum-energy wave functions.
If we look at Figures~\ref{truck} and \ref{parton33} we see that is
the effect of Lorentz covariance
described in these figures. One of the quark-parton
puzzles is thus resolved~\cite{knp86,kn77par,kim89}.

Another puzzling problem is that quarks are coherent when the proton
is at rest but the partons appear as incoherent particles.
We could ask whether this means that Lorentz boost coherence is destroyed.
Obviously, the answer to this question is NO. The resolution to this
puzzle is given below.

When the proton is boosted, its matter becomes squeezed. The result
is that the wave function for the proton becomes, along the positive
light-cone axis, concentrated in the elliptic region.
The major axis is expanded in length by
$\exp(\eta)$, and, as a consequence, the minor axis is contracted by
$\exp(-\eta)$.

Therefore we see that, among themselves, the interaction time of the
quarks becomes dilated. As~the wave function becomes wide-spread, the
ends of the harmonic oscillator well increase in distance from each
other. Universally observed in high-energy experiments, it was
Feynman~\cite{fey69a} who first observed this effect. The
oscillation period thus increase like $e^{\eta}$~\cite{bj69,hussar81}.

Since the external signal, on the other hand, is moving in the
direction opposite to the direction of the proton, it travels along
the negative light-cone axis with $t = -z$.
As the proton contracts along the negative light-cone axis,
the interaction time decreases by $\exp(-\eta)$.
Then the  ratio of the interaction time to the oscillator period becomes $\exp(-2\eta)$.
Each proton, produced by the Fermilab accelerator, used to have an
energy of $900~\rm{GeV}$. This then means the ratio is $10^{-6}$.
Because this is such small number, the external signal cannot sense,
inside the proton, the interaction of the quarks among themselves.



\subsection{Historical Note}\label{histo}

The hydrogen atom played a pivotal role in the development of
quantum mechanics.  Niels Bohr devoted much of his life to
understanding the electron orbit of the hydrogen atom.  Bohr
met Einstein occasionally to talk about physics.  Einstein's
main interest was how things look to moving observers.  Then,
did they discuss how the hydrogen atom looks to a moving
observer?

If they discussed this problem, there are no records.  If they
did not, they are excused.  At their time, the hydrogen atom
moving with a relativistic speed was not observable, and thus
beyond the limit of their scope.  Even these days, this atom
with total charge zero cannot be accelerated.

\begin{figure}
\centerline{\includegraphics[width=12cm]{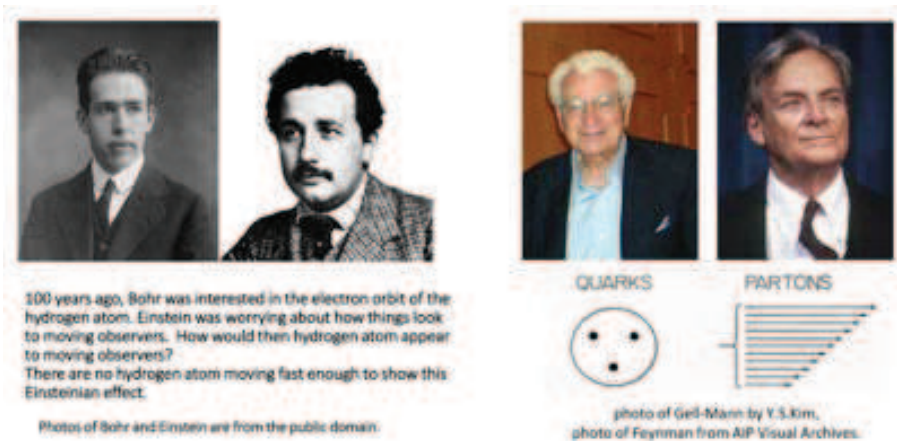}}
\vspace{5mm}
\caption{Bohr and Einstein, and then Gell-Mann and Feynman.
There are no records indicating that Bohr and Einstein discussed
how the hydrogen looks to moving observers.
After 1950, with particle accelerators, the physics world
started producing protons with relativistic speeds.
Furthermore, the~proton became a bound state sharing the same
quantum mechanics with the hydrogen atom.  The problem of
fast-moving hydrogen became
that of the proton.  How would the proton appear when it moves
with a speed close to that of light?  This is the quark-parton
puzzle.}\label{einbohr}
\end{figure}

After 1950, high-energy accelerators started producing protons
moving with relativistic speeds.  However, the proton is not
a hydrogen atom.  On the other hand, Hofstadter's
experiment~\cite{hofsta55} showed that the proton is not
a point particle.  In 1964, Gell-Mann produced the idea that
the proton is a bound-state of more fundamental particles called
{\em quarks}.  It is assumed that the proton and the hydrogen share
the same quantum mechanics in binding their constituent particles.
Thus, we can study the moving hydrogen atom by studying the
moving proton.  This historical approach is illustrated in
Figure~\ref{einbohr}.

\begin{figure}
\centerline{\includegraphics[width=12cm]{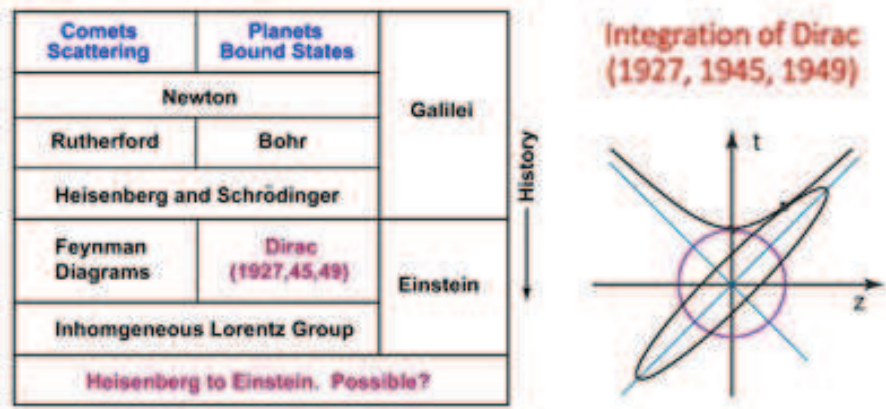}}
\vspace{5mm}
\caption{Scattering and bound states.  These days, Feynman diagrams
are used for scattering problems.  For bound-state problems, it
is possible to construct Lorentz-covariant harmonic oscillators
by integrating the papers written by Dirac.   Feynman diagrams and
the covariant oscillators are both two different representations
of the inhomogeneous Lorentz group.  Then is it possible to derive
Einstein's special relativity from the Heisenberg brackets?  This
problem is addressed in Sections~\ref{twopho}
and~\ref{contrac}.}\label{comet}
\end{figure}

Paul A. M. Dirac was interested in constructing localized wave
functions that can be Lorentz boosted.  He wrote three
papers~\cite{dir27,dir45,dir49}.  If we integrate his ideas,
it is possible to construct the covariant oscillator wave function
discussed in this paper. Figure~\ref{comet} tells where this
integration stands in the history of physics.

In constructing quantum mechanics in the Lorentz-covariant world,
the present form of quantum field theory is quite successful in
scattering problems where all participating particles are free in
the remote past and the remote future.  The bound states are different,
and Feynman suggested Lorentz-covariant harmonic oscillators
for studying this problem~\cite{fkr71}.  Yes, quantum field theory
and the covariant harmonic oscillators use quite different
mathematical forms. Yet, they share the same set of physical
principles~\cite{hkn81fp}.  They are both the representations of
the inhomogeneous Lorentz group.  We note that Dirac in his 1949
paper~\cite{dir49} said that we can build relativistic dynamics
by constructing representations of the inhomogeneous Lorentz group.
In Figure~\ref{comet}, both Feynman diagrams and the covariant
oscillator (integration of Dirac's papers) share the Lie
algebra of the inhomogeneous Lorentz group.  This figure leads
to the idea of whether the Lie algebra of quantum mechanics
can lead to that of the inhomogeneous Lorentz group. We shall
discuss this question in Section~\ref{contrac}.

Since the time of Bohr and Einstein, attempts have been made
to construct Lorentz-covariant bound states within the framework
of quantum mechanics and special relativity.  In recent years,
there~have been laudable efforts to construct a non-perturbative
approach to quantum field theory, where particles are in a bound
state and thus are not free in the remote past and remote
future~\cite{stro13}. If and when this approach  produces
localized probability distributions which can Lorentz-boosted,
it should explain both the quark model (at rest) and its parton
picture in the limit of large speed.

In recent years, there have been efforts to
represent the observation of movements
inside materials using Dirac electric states~\cite{park2020} as well
as to use relativistic methods to understand atomic and molecular
structure~\cite{grant2006}. There have also been efforts to provide
covariant formulation of the electrodynamics of nonlinear
media~\cite{hartemann}.

\subsection{Lorentz-Invariant Uncertainty Products}\label{uncert}
In the harmonic oscillator regime, the energy-momentum wave functions
take the same mathematical form, and the uncertainty relation in terms
of the uncertainty products is well understood.  However, in the present
case, the oscillator wave functions are deformed when Lorentz-boosted,
as~shown in Figure~\ref{parton33}.  According to this figure,  both the
space-time and momentum-energy wave functions become spread along their
longitudinal directions. Does this mean that the Lorentz boost increases
the uncertainty?

In order to address this question, let us write the momentum-energy
wave function as a Fourier transformation of the space-time wave
function:
\begin{equation}\label{fouri}
\phi\left(q_z, q_0\right) =
\frac{1}{2\pi} \int \psi(z, t)
             \exp\left(i\left[q_{z}z - q_{0} t\right]\right)~dt~dz .
\end{equation}
The transverse $x$ and $y$ components are not included in this
expression.  The exponent of this expression can be written as
\begin{equation}
 q_{z}z - q_{0} t = q_{+}z_{-} + q_{-}z_{+},
\end{equation}
 with
\begin{equation}
 q_{\pm} = \frac{1}{\sqrt{2}}\left(q_z \pm q_0 \right), \qquad
 z_{\pm} = \frac{1}{\sqrt{2}}\left(z \pm t \right) ,
\end{equation}
  as given earlier in Equations (\ref{lc11}) and (\ref{conju}).

In terms of these variables, the Fourier integral takes the form
\begin{equation}
\frac{1}{2\pi} \int \psi(z, t)
      \exp\left(i\left[q_{+}z_{-}  + q_{-}z_{+}\right]\right)~dt~dz .
\end{equation}
In this case, the variable
$q_{+}$ is conjugate to $z_{-}$, and $q_{- }$ to $z_{+}$.  Let us go
back to Figure~\ref{parton33}.  The major (minor) axis of the space-time
ellipse is conjugate to the minor (major) axis of the momentum-energy
ellipse. Thus the uncertainty products
\begin{equation}
    \left<z_{+}^{2}\right>  \left<q_{-}^{2}\right> \quad\mbox{and}\quad
    \left<z_{-}^{2}\right>  \left<q_{+}^{2}\right>
\end{equation}
remain invariant under the Lorentz boost.

\section{O(3,2) Symmetry Derivable from Two-Photon States}\label{twopho}

In this section we start with the paper Dirac published in 1963 on
the symmetries from two harmonic oscillators~\cite{dir63}.  Since
the step-up and step-down operators in the oscillator system are equivalent
to the creation and annihilation operators in the system of photons,
Dirac  was working with the system of two photons which is of current
interest~\cite{bkn19iop,dodo03,saleh07,walls08}.

In the oscillator system the step-up and step-down operators are:
\begin{eqnarray}\label{605}
&{}& a_{1} = \frac{1}{\sqrt{2}}\left(x_{1} + iP_{1}\right), \qquad
a_{1}^{\dag} = \frac{1}{\sqrt{2}}\left(x_{1} - iP_{1}\right),   \nonumber\\[2ex]
&{}& a_{2} = \frac{1}{\sqrt{2}}\left(x_{2} + iP_{2}\right), \qquad
a_{2}^{\dag} = \frac{1}{\sqrt{2}}\left(x_{2} - iP_{2}\right) ,
\end{eqnarray}
with
\begin{equation}
iP_{i} = \frac{\partial}{\partial x_{i}}.
\end{equation}

In terms of these operators, Heisenberg's uncertainty relations can be written as
\begin{equation}\label{607}
\left[a_{i}, a^{\dag}_{j}\right] = \delta_{ij} ,
\end{equation}
with
\begin{equation}\label{609}
  x_{i} = \frac{1}{\sqrt{2}}\left(a_{i} + a^{\dag}_{i} \right), \qquad
  P_{i} = \frac{i}{\sqrt{2}}\left(a^{\dag}_{i} - a_{i} \right).
\end{equation}

With these sets of operators, Dirac constructed three generators of the
form
\begin{equation}\label{611}
 J_{1} = {1\over 2}\left(a^{\dag}_{1}a_{2} + a^{\dag}_{2}a_{1}\right) ,
 \quad
 J_{2} = {1\over 2i}\left(a^{\dag}_{1}a_{2} - a^{\dag}_{2}a_{1}\right),
 \quad
 J_{3} = {1\over 2}\left(a^{\dag}_{1}a_{1} - a^{\dag}_{2}a_{2} \right),
\end{equation}
and three more of the form
\begin{eqnarray}\label{615}
&{}& K_{1} = -{1\over 4}\left(a^{\dag}_{1}a^{\dag}_{1} + a_{1}a_{1} -
  a^{\dag}_{2}a^{\dag}_{2} - a_{2}a_{2}\right) ,   \nonumber \\[2ex]
&{}& K_{2} = +{i\over 4}\left(a^{\dag}_{1}a^{\dag}_{1} - a_{1}a_{1} +
  a^{\dag}_{2}a^{\dag}_{2} - a_{2}a_{2}\right) ,    \\[2ex]
&{}& K_{3} = {1\over 2}\left(a^{\dag}_{1}a^{\dag}_{2} + a_{1}a_{2}\right) .\nonumber
\end{eqnarray}

These $J_{i}$ and $K_{i}$ operators satisfy the commutation relations
\begin{equation}\label{lie11}
 [J_{i}, J_{j}] = i\epsilon _{ijk} J_{k} ,\quad
 [J_{i}, K_{j}] = i\epsilon_{ijk} K_{k} ,  \quad
[K_{i}, K_{j}] = -i\epsilon _{ijk} J_{k} .
\end{equation}

This set of commutation relations is identical to the Lie algebra of the
Lorentz group where $J_{i}$ and $K_{i}$ are three rotation and three
boost generators respectively.   This set of commutators is the Lie
algebra of the Lorentz group with three rotation and three boost
generators.

In addition, with the harmonic oscillators, Dirac constructed another set
consisting of
\begin{eqnarray}
&{}& Q_{1} = -{i\over 4}\left(a^{\dag}_{1}a^{\dag}_{1} - a_{1}a_{1} -
  a^{\dag}_{2}a^{\dag}_{2} + a_{2}a_{2} \right) ,   \nonumber \\[2ex]
&{}& Q_{2} = -{1\over 4}\left(a^{\dag}_{1}a^{\dag}_{1} + a_{1}a_{1} +
   a^{\dag}_{2}a^{\dag}_{2} + a_{2}a_{2} \right), \\[3ex]
&{}& Q_{3} = \frac{i}{2}\left(a_{1}^{\dag}a_{2}^{\dag} - a_{1}a_{2}\right) .\nonumber
\end{eqnarray}

They then satisfy the commutation relations

\begin{equation}\label{lie22}
[J_{i}, Q_{j}] = i\epsilon_{ijk} Q_{k} , \quad
[Q_{i}, Q_{j}] = -i\epsilon _{ijk} J_{k} .
\end{equation}

Together with the relation $[J_{i}, J_{j}] = i\epsilon _{ijk} J_{k}$
given in Equation~(\ref{lie11}), $J_{i}$ and $Q_{i}$ produce another
Lie algebra of the Lorentz group.  Like $K_{i}$, the $Q_{i}$ operators
act as boost generators.

In order to construct a closed set of commutation relations for all the
generators, Dirac introduced an additional operator
 \begin{equation}\label{603}
S_{0} = {1\over 2}\left(a^{\dag}_{1}a_{1} + a_{2}a^{\dag}_{2}\right) .
\quad
\end{equation}

Then the commutation relations are
\begin{equation}\label{lie33}
  [K_{i}, Q_{j}] = -i\delta_{ij} S_{0} , \quad
[J_{i}, S_{0}] = 0 ,  \quad [K_{i}, S_{0}] =  -iQ_{i},\quad
[Q_{i}, S_{0}] = iK_{i} .
\end{equation}
It was then noted by Dirac that the three sets of commutation relations given
in Equations~(\ref{lie11}),~(\ref{lie22}) and~(\ref{lie33}) form the Lie
algebra for the $O(3,\,2)$ de Sitter group.  This group applies to
space of $(x,\, y,\, z,\, t,\, s)$, which is five-dimensional. In this
space, the three space-like coordinates are $x,\, y, \,z$,
and the time-like variables are given by $t$ and $s$.  Therefore,
these generators are five-by-five matrices.
The three rotation generators for the $(x,\, y, \,z)$ space-like
coordinates are given in Table~\ref{tab11}.  The three boost
generators with respect to the time variable $t$ are given in
Table~\ref{tab22}.  Table~\ref{tab33} contains three boost operators
with respect to the second time variable $s$ and the rotation
generator between the two time variables $t$ and $s$.

\begin{table}
\caption{Three generators of the rotations in the five-dimensional space of
$(x,\, y, \,z, \,t,\, s)$.  The time-like $s$ and $t$ coordinates are not affected
by the rotations in the three-dimensional space of $(x,\, y,\, z)$.}\label{tab11}
\begin{center}
\begin{tabular}{ccccccc}
\hline
\hline\\[-0.4ex]
& \textbf{Generators} & & \textbf{Differential}  &\hspace{8mm}& \textbf{Matrix}
\\
\hline\\
& $J_{1}$
& &  $ -i\left(y\frac{\partial}{\partial z} - z\frac{\partial}{\partial y}\right) $
 & \hspace{8mm}&
$\pmatrix{0 & 0 & 0 & 0  & 0 \cr 0 & 0 & -i & 0 & 0 \cr
  0 & i & 0 & 0 & 0 \cr 0 & 0 & 0 & 0 & 0 \cr 0 & 0 & 0 & 0 & 0  } $\\
\hline\\
&
$J_{2} $
& &
$  -i\left(z\frac{\partial}{\partial x} - x\frac{\partial}{\partial z}\right)$
                &  \hspace{8mm} &
$ \pmatrix{0 & 0 & i & 0 & 0 \cr 0 & 0 & 0 & 0 & 0 \cr
-i & 0 & 0 & 0 & 0 \cr 0 & 0 & 0 & 0 & 0 \cr 0 & 0 & 0 & 0 & 0 } $\\
\hline\\
&
$J_{3} $ & &
$ -i\left(x\frac{\partial}{\partial y} - y\frac{\partial}{\partial x}\right)$
                  & \hspace{8mm}&
$ \pmatrix{0 & -i & 0 & 0 & 0 \cr i & 0 & 0 & 0 & 0 \cr
0 & 0 & 0 & 0 & 0 \cr 0 & 0 & 0 & 0 & 0 \cr\    0 & 0 & 0 & 0 & 0 }  $\\
\hline
\hline\\[-0.4ex]
\end{tabular}
\end{center}
\end{table}


\begin{table}
\caption{Three generators of Lorentz boosts with respect the time variable
$t$.   The $s$ coordinate is not affected by these boosts. }\label{tab22}
\begin{center}
\begin{tabular}{ccccccc}
\hline
\hline\\[-0.4ex]
& \textbf{Generators} & & \textbf{Differential}  &\hspace{8mm}& \textbf{Matrix}\\
\hline\\
& $K_{1} $ & &
$ -i\left(x\frac{\partial}{\partial t} + t\frac{\partial}{\partial x}\right)$
                  & \hspace{8mm}&
$ \pmatrix{0 & 0 & 0 & i & 0 \cr 0 & 0 & 0 & 0 & 0 \cr
0 & 0 & 0 & 0 & 0 \cr i & 0 & 0 & 0 & 0 \cr    0 & 0 & 0 & 0 & 0 }  $\\
\hline\\
& $K_{2} $ & &
$ -i\left(y\frac{\partial}{\partial t} + t\frac{\partial}{\partial y}\right)$
                  & \hspace{8mm}&
$ \pmatrix{0 & 0 & 0 & 0 & 0 \cr 0 & 0 & 0 & i & 0 \cr
0 & 0 & 0 & 0 & 0 \cr 0 & i & 0 & 0 & 0 \cr 0 & 0 & 0 & 0 & 0 }  $\\
\hline\\
& $K_{3} $ & &
$ -i\left(z\frac{\partial}{\partial t} + t\frac{\partial}{\partial z}\right)$
                  & \hspace{8mm}&
$ \pmatrix{0 & 0 & 0 & 0 & 0 \cr 0 & 0 & 0 & 0 & 0 \cr
0 & 0 & 0 & i & 0 \cr 0 & 0 & i & 0 & 0 \cr    0 & 0 & 0 & 0 & 0  } $\\
\hline
\hline\\[-0.4ex]
\end{tabular}
\end{center}
\end{table}
\begin{table}
\caption{The $O(3,\,2)$ group has four additional generators. Note
that the generators in this table
have non-zero elements only in the fifth row and the fifth
column. This is unlike those given in Tables~\ref{tab11} and \ref{tab22}.
Here the $s$ variable is contained in every differential operator. }\label{tab33}
\vspace{0.5mm}
\begin{center}
\begin{tabular}{ccccccc}
\hline
\hline\\[-0.4ex]
& \textbf{Generators} & & \textbf{Differential}  &\hspace{8mm}& \textbf{Matrix}\\
\hline\\
& $Q_{1}$
& &  $ -i\left(x\frac{\partial}{\partial s} +
     s\frac{\partial}{\partial x}\right) $
 & \hspace{8mm}&
$\pmatrix{0 & 0 & 0 & 0  & i \cr 0 & 0 & 0 & 0 & 0 \cr
  0 & 0 & 0 & 0 & 0 \cr 0 & 0 & 0 & 0 & 0 \cr i & 0 & 0 & 0 & 0 } $
\\
\hline\\
&
$Q_{2} $
& &
$  -i\left(y\frac{\partial}{\partial s} + s\frac{\partial}{\partial y}\right)$
                &  \hspace{8mm} &
$ \pmatrix{0 & 0 & 0 & 0 & 0 \cr 0 & 0 & 0 & 0 & i \cr
0 & 0 & 0 & 0 & 0 \cr 0 & 0 & 0 & 0 & 0 \cr 0 & i & 0 & 0 & 0 }  $ \\
\hline\\
& $Q_{3} $ & &
$ -i\left(z\frac{\partial}{\partial s} + s\frac{\partial}{\partial z}\right)$
                  & \hspace{8mm}&
$ \pmatrix{0 & 0 & 0 & 0 & 0 \cr 0 & 0 & 0 & 0 & 0 \cr
0 & 0 & 0 & 0 & i \cr 0 & 0 & 0 & 0 & 0 \cr  0 & 0 & i & 0 & 0  } $ \\
\hline\\
& $S_{0} $ & &
$ -i\left(t\frac{\partial}{\partial s} - s\frac{\partial}{\partial t}\right)$
                  & \hspace{8mm}&
$ \pmatrix{0 & 0 & 0 & 0 & 0 \cr 0 & 0 & 0 & 0 & 0 \cr
0 & 0 & 0 & 0 & 0 \cr 0 & 0 & 0 & 0 & -i \cr  0 & 0 & 0 & i & 0  }  $ \\
\hline
\hline\\[-0.4ex]
\end{tabular}
\end{center}
\end{table}

It is indeed remarkable, as Dirac stated in his paper~\cite{dir63},
that the space-time symmetry of the \mbox{(3 + 2)} de Sitter group is the
result of this two-oscillator system.  What is even more remarkable
is that we can derive, from quantum optics, this two-oscillator system.
In the two-photon system in optics, where $i$ can be 1 or 2,
$a_{i}$ and $a^{\dag}_{i}$ act as the annihilation and
creation operators.

It is possible to construct, with these two sets of operators,
two-photon states~\cite{bkn19iop}.  Yuen, as early as 1976~\cite{yuen76},
used the two-photon state generated by
\begin{equation}\label{301a}
 Q_{3} = \frac{i}{2}\left(a_{1}^{\dag}a_{2}^{\dag} - a_{1}a_{2}\right).
\end{equation}
This leads to the two-mode coherent state. This is also known as the
{\em squeezed state}.

It was later that Yurke, McCall, and Klauder, in 1986~\cite{yurke86},
investigated two-mode interferometers.  In their study of two-mode states, they
started with $Q_{3}$ given in Equation~(\ref{301a}).  Then they considered
that the following two additional operators,
\begin{equation}\label{303}
K_{3} = {1\over 2}\left(a^{\dag}_{1}a^{\dag}_{2} + a_{1}a_{2}\right) ,
\qquad
S_{0} = {1\over 2}\left(a^{\dag}_{1}a_{1} + a_{2}a^{\dag}_{2}\right) ,
\end{equation}
were needed in one of their interferometers.
These three Hermitian operators from Equations~(\ref{301a}) and (\ref{303})
have the following commutation relations
\begin{equation} \label{305}
\left[K_{3}, Q_{3}\right] = -iS_{0}, \qquad
\left[Q_{3}, S_{0}\right] = iK_{3}, \qquad
\left[S_{0}, K_{3}\right] = iQ_{3} .
\end{equation}
Yurke et al. called this device the $SU(1,\,1)$ interferometer.
The group $SU(1,\,1)$ is isomorphic to the $O(2,\,1)$ group or the
Lorentz group applicable to two space-like and one time-like
dimensions.

In addition, in the same paper~\cite{yurke86}, Yurke et al.
discussed the possibility of constructing another interferometer
exhibiting the symmetry generated by
\begin{equation}\label{307}
 J_{1} = {1\over 2}\left(a^{\dag}_{1}a_{2} + a^{\dag}_{2}a_{1}\right) , \quad
 J_{2} = {1\over 2i}\left(a^{\dag}_{1}a_{2} - a^{\dag}_{2}a_{1}\right), \quad
 J_{3} = {1\over 2}\left(a^{\dag}_{1}a_{1} - a^{\dag}_{2}a_{2} \right).
\end{equation}
These generators satisfy the closed set of commutation relations
\begin{equation}
\left[J_{i}, J_{j}\right] = i\epsilon_{ijk} J_{k} ,
\end{equation}
given in Equation~(\ref{611}).  This is the Lie algebra for the three-dimensional
rotation group.  Yurke et al. called this optical device the $SU(2)$
interferometer.

We are then led to ask whether it is possible to construct a closed set
of commutation relations with the six Hermitian operators from
Equations~(\ref{305}) and (\ref{307}). It is not possible.  We have to
add four additional operators, namely
\begin{eqnarray}\label{311}
&{}& K_{1} = -{1\over 4}\left(a^{\dag}_{1}a^{\dag}_{1} + a_{1}a_{1} -
  a^{\dag}_{2}a^{\dag}_{2} - a_{2}a_{2}\right) ,    \nonumber \\[1ex]
&{}&   K_{2} = +{i\over 4}\left(a^{\dag}_{1}a^{\dag}_{1} - a_{1}a_{1} +
  a^{\dag}_{2}a^{\dag}_{2} - a_{2}a_{2}\right) ,   \\[1ex]
&{}& Q_{1} = -{i\over 4}\left(a^{\dag}_{1}a^{\dag}_{1} - a_{1}a_{1} -
  a^{\dag}_{2}a^{\dag}_{2} + a_{2}a_{2} \right) ,    \nonumber \\[1ex]
&{}& Q_{2} = -{1\over 4}\left(a^{\dag}_{1}a^{\dag}_{1} + a_{1}a_{1} +
   a^{\dag}_{2}a^{\dag}_{2} + a_{2}a_{2} \right) . \nonumber
\end{eqnarray}

There are now ten operators.  They are precisely those ten Dirac
constructed in his paper of 1963~\cite{dir63}.

It is indeed remarkable that Dirac's $O(3,\,2)$ algebra is produced by
modern optics. This algebra produces  the Lorentz group applicable
to three space-like and two time-like dimensions.

The algebra of harmonic oscillators given in Equation~(\ref{607}) is
Heisenberg's uncertainty relations in a two-dimensional space.  The
de Sitter group $O(3,\,2)$ is basically a language of
special relativity.   Does this mean that Einstein's special relativity
can be derived from the Heisenberg brackets?  We shall examine this
problem in Section~\ref{contrac}.

\section{Contraction of  O(3, \,2) to the Inhomogeneous
Lorentz Group}\label{contrac}

According to Section~\ref{twopho},  the group $O(3,\,2)$ has an $O(3,\,1)$
subgroup with six generators plus four additional generators.  The
inhomogeneous Lorentz group also contains one Lorentz group $O(3,\,1)$
as its subgroup plus four space-time translation generators.  The
question arises whether the four generators in Table~\ref{tab33} can
be converted into the four translation generators.   The purpose of
this section is to prove this is possible according to the group
contraction procedure introduced first by In{\"o}n{\"u} and
Wigner~\cite{inonu53}.

In their paper In{\"o}n{\"u} and Wigner introduced the procedure for
transforming the Lorentz group into the Galilei group.  This procedure
is known as {\em group contraction}~\cite{inonu53}.  In this section,
we use the same procedure to contract the $O(3,\,2)$ group into the
inhomogeneous Lorentz group which is the group $O(3,\,1)$ plus four
translations.

\begin{table}
\caption{Here the generators of translations are given in the four-dimensional Minkowski
space. It~is of interest to convert the four generators in
the $O(3,\,2)$ group in Table~\ref{tab33}
into the four translation~generators.}\label{tab44}
\vspace{0.5mm}
\begin{center}
\begin{tabular}{ccccccc}
\hline
\hline\\[-0.4ex]
& \textbf{Generators} & & \textbf{Differential}  &\hspace{8mm}& \textbf{Matrix}\\
\hline\\
&
$Q_{1}\rightarrow P_{1}$ & &  $ -i\frac{\partial}{\partial x} $
 & \hspace{8mm}&
$\pmatrix{0 & 0 & 0 & 0  & i \cr 0 & 0 & 0 & 0 & 0 \cr
  0 & 0 & 0 & 0 & 0 \cr 0 & 0 & 0 & 0 & 0 \cr 0 & 0 & 0 & 0 & 0   } $ \\
\hline\\
&
$Q_{2}\rightarrow P_{2} $  & &
$  -i\frac{\partial}{\partial y}$  &  \hspace{8mm} &
$ \pmatrix{0 & 0 & 0 & 0 & 0 \cr 0 & 0 & 0 & 0 & i \cr
0 & 0 & 0 & 0 & 0 \cr 0 & 0 & 0 & 0 & 0 \cr 0 & 0 & 0 & 0 & 0  } $ \\
\hline\\
&
$Q_{3} \rightarrow P_{3} $ & & $ -i\frac{\partial}{\partial z}$ & \hspace{8mm}&
$ \pmatrix{0 & 0 & 0 & 0 & 0 \cr 0 & 0 & 0 & 0 & 0 \cr
0 & 0 & 0 & 0 & i \cr 0 & 0 & 0 & 0 & 0 \cr  0 & 0 & 0 & 0 & 0  } $\\
\hline\\
&
$S_{0}\rightarrow P_{0} $ & & $ i \frac{\partial}{\partial t}$ & \hspace{8mm}&
$ \pmatrix{0 & 0 & 0 & 0 & 0 \cr 0 & 0 & 0 & 0 & 0 \cr
0 & 0 & 0 & 0 & 0 \cr 0 & 0 & 0 & 0 & -i \cr  0 & 0 & 0 & 0 & 0 } $ \\
\hline
\hline\\[-0.4ex]
\end{tabular}
\end{center}
\end{table}

Let us introduce the contraction matrix
\begin{equation}
  C = \pmatrix{ 1/\epsilon & 0 & 0& 0 & 0 \cr
0 &1/\epsilon & 0 & 0 & 0 \cr 0 & 0 & 1/\epsilon & 0 & 0 \cr
0 & 0 & 0 &1/\epsilon & 0\cr 0 & 0 & 0 & 0 & \epsilon   }.
\end{equation}

This matrix expands the $z, \,y,\, z,\, t$ axes, and contracts $s$ axis as $\epsilon$
becomes small.  Yet, the $J_{i}$ and $K_{i}$ matrices given in Tables~\ref{tab11}
and~\ref{tab22} remain invariant:
\begin{equation}
  CJ_{i} C^{-1} = J_{i}, \quad\mbox{and}\quad  C K_{i} C^{-1} = K_{i}.
\end{equation}

These matrices have zero elements in the fifth row and fifth column.

On the other hand, the matrices in Table~\ref{tab33} are different.  Let us
choose $Q_{3}$.  The same algebra leads to the elements $\epsilon^2$ and
its inverse, as shown in this equation:
\begin{eqnarray}\label{707}
 C Q_{3}C^{-1}& =& \pmatrix{ 1/\epsilon & 0 & 0& 0 & 0 \cr
0 & 1/\epsilon & 0 & 0 & 0 \cr 0 & 0 & 1/\epsilon & 0 & 0 \cr
0 & 0 & 0 &1/\epsilon & 0\cr 0 & 0 & 0 & 0 & \epsilon   }
\pmatrix{0 & 0 & 0 & 0 & 0 \cr 0 & 0 & 0 & 0 & 0 \cr
0 & 0 & 0 & 0 & i  \cr 0 & 0 & 0 & 0 & 0 \cr  0 & 0 & i & 0 & 0  }
\pmatrix{ \epsilon & 0 & 0& 0 & 0 \cr
0 & \epsilon & 0 & 0 & 0 \cr 0 & 0 & \epsilon & 0 & 0 \cr
0 & 0 & 0 & \epsilon & 0\cr 0 & 0 & 0 & 0 & 1/ \epsilon   } \\[4ex]
 &=&
 \pmatrix{ 0 & 0 & 0& 0 & 0 \cr 0 & 0 & 0 & 0 & 0 \cr
0 & 0 & 0 & 0 & i/\epsilon^2  \cr 0 & 0 &  0 & 0 & 0 \cr
0 & 0 & 0 & i \epsilon^2 & 0 } \rightarrow
 P_{3} = \pmatrix{ 0 & 0 & 0& 0 & 0 \cr 0 & 0 & 0 & 0 & 0 \cr
0 & 0 & 0 & 0 & i \cr 0 & 0 &  0 & 0 & 0 \cr 0 & 0 & 0 & 0 & 0 } .
\end{eqnarray}

The second line of this equation tells us that $\epsilon^2$ becomes zero in
the limit of small $\epsilon$.  If we make the inverse transformation,
result is $P_{3}$.  We can perform the same algebra to arrive at the
the results given in Table~\ref{tab44}.  According to this table, $Q_{i}$ and
$S_{0}$ become contracted to the generators of the space-time translations.
In other words, the de Sitter group $O(3,\,2)$ can be contracted to the
inhomogeneous Lorentz group.

\begin{figure}
\centerline{\includegraphics[width=11cm]{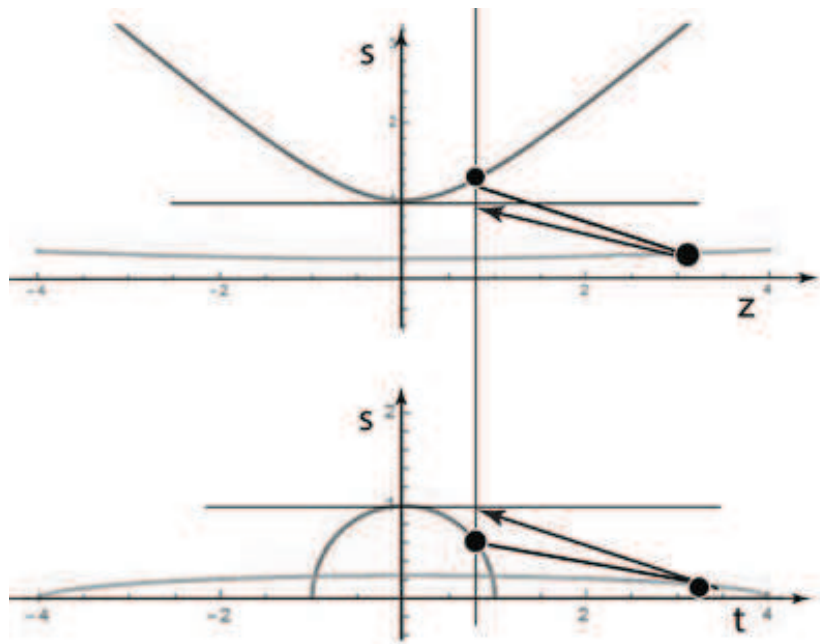}}
\vspace{5mm}
\caption{Contraction of $O(3,\, 2)$ to the inhomogeneous Lorentz group.
The extra time variable $s$ becomes a constant, as shown by a flat line
in this figure.}\label{o32po}
\end{figure}

If this matrix is applied to the five-vector of $(x, \, y,\,z,\, t,\, s)$, it
becomes $(x/\epsilon, \, y/\epsilon, \, z/\epsilon, \, t/\epsilon, \, s \epsilon)$.
If~$\epsilon$ becomes very small, the $s$ axis contracts while the
four others expand. As $\epsilon$ becomes very small, $\epsilon s$
approaches zero and can be replaced by $\epsilon$, since both of them
are zero.  We are replacing zero by another zero, as shown here:
\begin{equation}
   \pmatrix{x/\epsilon \cr y/\epsilon \cr z/\epsilon \cr
    t/\epsilon \cr
\epsilon s }  \rightarrow
\pmatrix{x/\epsilon \cr y/\epsilon \cr z/\epsilon \cr
   t/\epsilon \cr \epsilon } ,
\quad\mbox{and}\quad
 \pmatrix{ \epsilon & 0 & 0& 0 & 0 \cr
0 & \epsilon & 0 & 0 & 0 \cr 0 & 0 & \epsilon & 0 & 0 \cr
0 & 0 & 0 & \epsilon & 0\cr 0 & 0 & 0 & 0 & 1/\epsilon  }
\pmatrix{x/\epsilon \cr y/\epsilon \cr z/\epsilon \cr
   t/\epsilon \cr \epsilon } =
   \pmatrix{ x \cr y \cr z \cr t \cr 1}.
\end{equation}
This contraction procedure is indicated in Figure~\ref{o32po}.

Indeed, the five-vector $(x,\, y,\, z,\, t,\, 1)$ serves as the space-time
five-vector of the inhomogeneous Lorentz group.  The transformation
matrix applicable to this five-vector consists of that of the Lorentz
group in its first four row and columns.  Let us see how the
generators $P_{i}$ and $S_{0}$ generate translations.  First of all,
in terms of these generators, the transformation matrix takes the
form
\begin{equation}\label{770}
\exp\left(-i\left[a P_{x} + b P_{y} + z P_{z} + d P_{t} \right]\right)
 = \pmatrix{1 & 0 & 0 & 0 & a \cr 0 & 1 & 0 & 0 & b \cr
0 & 0 & 1 & 0 & c \cr 0 & 0 & 0 & 1 & d \cr 0 & 0 & 0 & 0 & 1 }.
\end{equation}

If this matrix is applied to the five-vector $(x,\, y,\, z,\, t,\, 1)$, the
result is the translation:
\begin{equation}
\pmatrix{1 & 0 & 0 & 0 & a \cr 0 & 1 & 0 & 0 & b \cr
0 & 0 & 1 & 0 & c \cr 0 & 0 & 0 & 1 & d \cr 0 & 0 & 0 & 0 & 1 }
\pmatrix{ x \cr y \cr z \cr t \cr 1 } =
\pmatrix{ x + a \cr y + b \cr z + c \cr t + d \cr 1}.
\end{equation}

The five-by-five matrix of Equation~(\ref{770}) indeed performs the translations.

Let us go to Table~\ref{tab44} again.  The four differential forms of the
translation generators correspond to the four-momentum satisfying the
equation $E^2 = p_{x}^2 + p_{y}^2 + p_{z}^2 + m^2$.  This is of course
Einstein's $E = mc^{2}$.

\section{Concluding Remarks}

Since 1973, the present authors have been publishing papers on the
harmonic oscillator wave functions which can be Lorentz-boosted.
The covariant harmonic oscillator plays roles in understanding
some of the properties in high-energy particle physics. This
covariant oscillator can also be used as a representation of
Wigner's little group for massive particles.

More recently, the oscillator wave function was shown to provide basic
mathematical tools for two-photon coherent states known as the squeezed
state of light~\cite{bkn19iop}.

In this paper, we have provided a review of our past efforts, with the
purpose of integrating the papers Paul A. M. Dirac wrote in his lifelong
efforts to make quantum mechanics compatible with Einstein's special
relativity which produced the Lorentz-covariant energy-momentum relation.
It is interesting to note that Einstein's special relativity is derivable
from Heisenberg's expression of the uncertainty relation.

\end{document}